\documentclass[preprint2,times,tighten]{aastex6}
\usepackage{amsmath,amstext}
\usepackage[all]{hypcap} 

\shorttitle{Raman scattering in exoplanet atmospheres}
\shortauthors{Oklop\v{c}i\'c, Hirata \& Heng}

\begin{document}

\title{Raman scattering by molecular hydrogen and nitrogen in exoplanetary atmospheres}
\author{Antonija Oklop\v ci\'c\altaffilmark{1}, Christopher M. Hirata\altaffilmark{2} and Kevin Heng\altaffilmark{3}}
\email{oklopcic@astro.caltech.edu}
\altaffiltext{1}{California Institute of Technology, MC 249-17, 1200 East California Boulevard, Pasadena, California 91125 }
\altaffiltext{2}{Center for Cosmology and Astroparticle Physics, Ohio State University, 191 West Woodruff Avenue, Columbus, Ohio 43210 }
\altaffiltext{3}{Center for Space and Habitability, University of Bern, Sidlerstrasse 5, CH-3012, Bern, Switzerland}

\begin{abstract}
An important source of opacity in exoplanet atmospheres at short visible and near-UV wavelengths is Rayleigh scattering of light on molecules. It is accompanied by a related, albeit weaker process -- Raman scattering. We analyze the signatures of Raman scattering imprinted in the reflected light and the geometric albedo of exoplanets, which could provide information about atmospheric properties. Raman scattering affects the geometric albedo spectra of planets in following ways. Firstly, it causes filling-in of strong absorption lines in the incident radiation, thus producing sharp peaks in the albedo. Secondly, it shifts the wavelengths of spectral features in the reflected light causing the so-called Raman ghost lines. Raman scattering can also cause a broadband reduction of the albedo due to wavelength shifting of a stellar spectrum with red spectral index. Observing the Raman peaks in the albedo could be used to measure the column density of gas, thus providing constrains on the presence of clouds in the atmosphere. Observing the Raman ghost lines could be used to spectroscopically identify the main scatterer in the atmosphere, even molecules like H$_2$ or N$_2$ that do not have prominent spectral signatures in the optical wavelength range. If detected, ghost lines could also provide information about the temperature of the atmosphere. In this paper we investigate the effects of Raman scattering in hydrogen- and nitrogen-dominated atmospheres. We analyze the feasibility of detecting the signatures of Raman scattering with the existing and future observational facilities, and of using these signatures as probes of exoplanetary atmospheres.
\end{abstract}

\maketitle

\section{Introduction}
Characterizing atmospheres of exoplanets usually involves detecting signatures of strong absorbers present in the atmosphere, such as alkali atoms sodium and potassium, and molecules like H$_2$O, CH$_4$, CO and CO$_2$ \citep[e.g.][]{SeagerSasselov2000, Brown2001, Hubbard2001, Sudarsky2003}. However, these atomic and molecular species can often be present in small amounts and the bulk of the atmosphere can be made of spectroscopically inactive species -- molecular hydrogen and helium in the case of giant planets, or molecular nitrogen in the case of the Earth's atmosphere. Determining the composition of the spectroscopically inactive gas components is very important for atmosphere characterization. Constraining the abundance of bulk gases breaks the degeneracies associated with measuring the (relative) abundances of trace gases and other atmospheric properties, such as the atmospheric pressure and temperature which can affect the strength of absorption lines. However, detecting molecules like H$_2$ and N$_2$ directly (i.e. spectroscopically) is challenging because these are homonuclear diatomic molecules with no low-lying excited electronic states, and hence do not have prominent spectral features in the optical wavelength range. 

At high densities, these molecules can form temporary collisional pairs that have non-zero electric dipole moments and therefore exhibit radiative dipole transitions. Via this so-called collision-induced absorption (CIA) mechanism, H$_2$-H$_2$ and H$_2$-He pairs become an additional source of opacity in the infrared part of the spectrum \citep{Saumon2012} that can be used to constrain the abundance of molecular hydrogen and helium. \cite{Schwieterman2015} similarly proposed using an absorption feature of a collisional pair of N$_2$ molecules at 4.3 $\mu$m to detect and constrain the abundance of molecular nitrogen in the atmospheres of terrestrial exoplanets. These broad spectral features of H$_2$ and N$_2$ can be very useful probes of deep and dense atmospheres. However, the spectral range in which they reside (NIR and IR) contains numerous spectral lines/bands of other strong absorbers like methane, water, and carbon dioxide with which CIA features often overlap, which can lead to uncertainties in the measured abundances. Therefore, having a complementary method of detecting signatures of H$_2$ or N$_2$ in the visible part of the spectrum would be very useful for breaking some of these degeneracies.

An important source of opacity at short optical and UV wavelengths is Rayleigh scattering of radiation by abundant molecules like H$_2$ \citep{Marley1999, Sudarsky2000, Hubbard2001}. Rayleigh scattering has been observed in transmission spectra of hot Jupiters \citep{Lecavelier2008a, Lecavelier2008b}, however this is not an unambiguous sign of H$_2$ because the observed spectral slope could also be produced by Rayleigh scattering by condensate particles, as discussed by \cite{Lecavelier2008a}. Rayleigh scattering does not leave direct spectroscopic signatures such as distinct spectral lines, that could be used to unambiguously identify the main scatterer. 

A related, albeit weaker, process -- Raman scattering -- does imprint signatures of scattering molecules in the scattered light and hence in the albedo of planets. If these Raman features in the albedo were detected, they could be used to distinguish between molecular scattering and scattering by small particles which does not produce Raman features, hence providing information on the presence of clouds or hazes in the observed atmosphere. Additionally, Raman scattering features might be used to spectroscopically identify the dominant scattering molecule, such as H$_2$ or N$_2$, and even the relative population of different molecular levels, thus constraining the temperature of the medium. 

The effects of Raman scattering on planetary albedos and its potential for probing planetary atmospheres has been known and studied for a few decades, mostly in the context of Solar System planets \citep[for reviews of the early studies of Raman scattering see][]{Price1977, Cochran1981a}. The purpose of this paper is to show that detecting the signatures of Raman scattering in the reflected light from \textit{exoplanets} could be a powerful tool in exoplanet atmosphere characterization and hence may become an interesting target for observations in the era of the next-generation observational facilities. In this paper we investigate the effects of Raman scattering by molecular hydrogen and nitrogen on the geometric albedo spectra of exoplanets and analyze the feasibility of detecting these features in nearby exoplanets using the currently available and future observational facilities. 

In Section~\ref{sec:raman} we give an overview of the basic physics related to Raman scattering and describe the effects of Raman scattering on the reflected light from planets and their albedo. In Section~\ref{sec:methods} we describe the radiative transfer calculations and model atmospheres that we use in this study. We present the results of radiative transfer calculations for two test cases -- a monochromatic light source and a flat photon spectrum with an absorption line in Section~\ref{sec:results}. The goal of these test cases is to provide insight into the way in which Raman scattering affects the reflected light from planets. We then present the calculated albedo spectra for several models of clear and cloudy hydrogen- and nitrogen-dominated atmospheres irradiated by the solar spectrum. We discuss our results and analyze the detectability of Raman features in the reflected light from exoplanets in Section~\ref{sec:discussion}. In Section~\ref{sec:conclusion} we present the summary and the conclusions of this work.

\section{Raman scattering}
\label{sec:raman}

\begin{figure}
\centering
\includegraphics[width=0.48\textwidth]{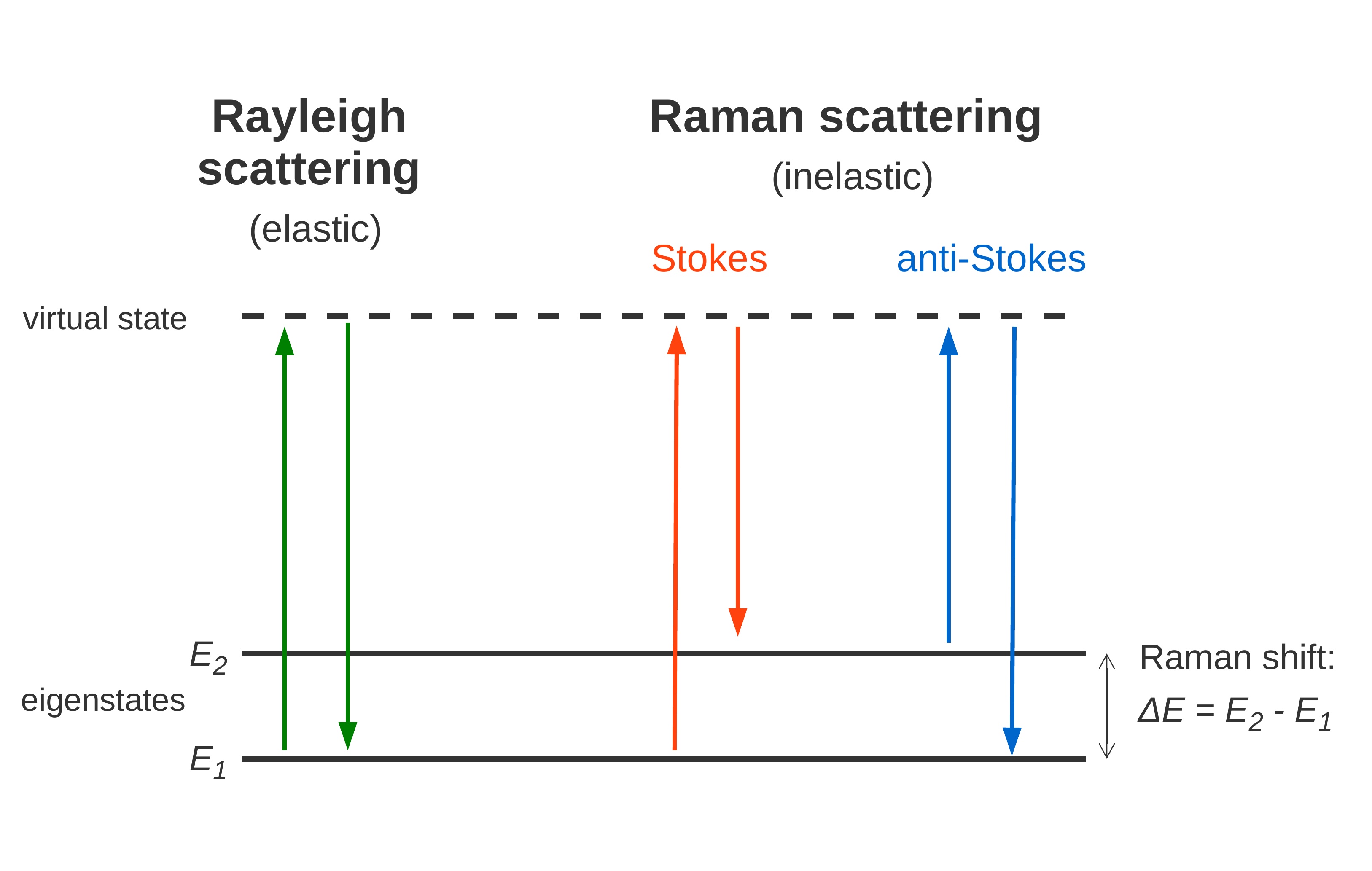}
\caption{Schematic representation of Rayleigh and Raman scattering. The initial and final states are eigenstates of the molecule, whereas the intermediate state is virtual, making these processes sources of continuum opacity. The energy of the final state can be higher (Stokes Raman scattering), lower (anti-Stokes scattering) or equal to that of the initial state (Rayleigh scattering). The change in the photon energy (Raman shift) is determined by the structure of molecular energy levels, and is unique for every molecule.}
\label{fig:scheme}
\end{figure}

\begin{figure*}
\centering
\includegraphics[height=0.32\textheight]{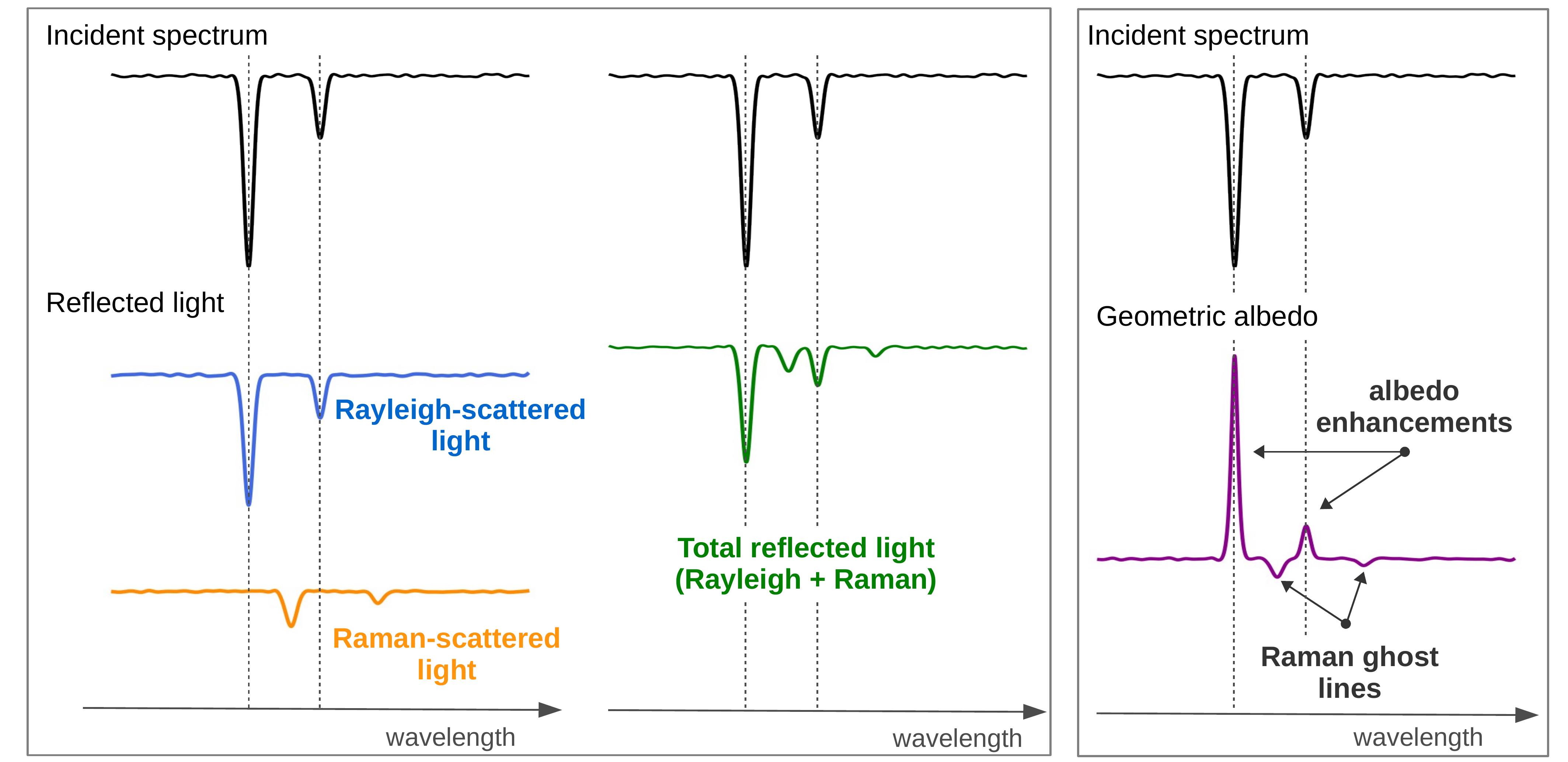}
\caption{Reflected light due to Rayleigh scattering is proportional to the stellar flux at the same wavelength. The contribution from Raman scattering depends on the incident flux at the Raman-shifted wavelengths. If there are strong features such as absorption lines in the incident spectrum, they will be present in the reflected light at the same wavelength (due to Rayleigh scattering) and at shifted positions (due to Raman scattering). Consequently, there are two types of features in the albedo spectrum due to Raman scattering: i) albedo enhancements at the wavelengths of absorption lines due to filling-in of the lines by Raman-shifted light and ii) dips in the albedo at wavelengths of the so-called Raman ghosts of the absorption lines. A pure Rayleigh-scattering atmosphere (with no Raman scattering and no absorption) would result in a flat, featureless geometric albedo spectrum.}
\label{fig:cartoon}
\end{figure*}

Scattering is an interaction between radiation and matter in which an incident photon is annihilated, and a new photon is created. If the scattered photon has the same energy as the incident one, the process is called Rayleigh scattering. If the energies of the incident and the scattered photon are different, we talk about Raman scattering. Raman and Rayleigh scattering are sources of continuum opacity because in both cases the intermediate state of the atom or the molecule is a virtual state, not an eigenstate of the system, as in the case of resonance scattering. For a detailed treatment of Raman scattering we point the reader to \cite{Long2002}.

Fig. \ref{fig:scheme} shows schematically the mechanism of Raman scattering, as well as Rayleigh scattering which may be thought of as a special case of Raman scattering. Depending on whether the scattered photon has lower or higher energy than the incident one, we distinguish between Stokes and anti-Stokes Raman processes. Due to Raman scattering the molecule changes its rotational and/or vibrational state. The allowed changes in the rotational quantum number of the molecule are $\Delta J\equiv J_f-J_i = -2, 0, 2$, corresponding to transitions with spectroscopic labels O, Q, and S, respectively. Vibrational quantum number of the molecule changes in the case of pure vibrational (Q-branch) or rotational-vibrational (O- and S-branch) Raman scattering. Raman transitions depend on the molecular polarizability, not on its electric dipole moment like the regular rotational and vibrational transitions. That is why even homonuclear diatomic molecules, like H$_2$ and N$_2$, that have zero dipole moments and hence do not exhibit rotational and ro-vibrational transitions, do exhibit Raman transitions, since their polarizability is non-zero.

The change in the energy of the scattered photon, i.e. the Raman shift, is determined by the energy difference between the initial and the final state of the molecule. This property of Raman scattering can be used to spectroscopically identify the scattering medium by measuring the Raman shift that photons experience. Because the exact value of the Raman shift depends on the initial and the final state of the molecule and because the intensity of a shifted line is proportional to the population of the initial state (in the optically thin case), a detailed analysis of the Raman scattering features and their relative intensities can provide information about the relative population of different molecular levels, and hence be used to constrain the temperature of the scattering medium.

If the intensity of light reflected off a planet is directly proportional to the intensity of the incident stellar light at the same wavelength, as it is in the case of a deep (semi-infinite) Rayleigh-scattering atmosphere without any absorption, the resulting geometric albedo spectrum of the planet (the ratio of the reflected and incident light as a function of wavelength) is flat and featureless.\footnote{For a Rayleigh-scattering atmosphere of finite depth the albedo spectrum is not necessarily flat, as shown in \citet[Fig. 2]{Sromovsky2005b}, but it is always a smoothly-varying function of optical depth, and consequently, of frequency. Hence, sharp features like spectral lines do not appear in the albedo spectrum caused by Rayleigh scattering.} On the other hand, the intensity of Raman-scattered light at any wavelength depends on the intensity of the incident radiation at a different, Raman-shifted wavelength. Due to this additional component, the total reflected spectrum (comprising of Rayleigh and Raman contributions) is not simply proportional to the incident spectrum and the resulting albedo is not smooth -- it contains spectral features characteristic of Raman scattering. In this work, we focus on two types of features that appear in the planetary albedo spectrum as a result of Raman scattering (see Fig. \ref{fig:cartoon} for reference): 
\begin{enumerate}
\item \textbf{Albedo enhancements}: If the incident spectrum has some distinct spectral features, such as strong absorption lines, they will be present in the reflected light at the same wavelength due to Rayleigh scattering. However, the Raman-scattered component at that wavelength is related to the intensity of the incident radiation at the Raman-shifted wavelength, where there is (most likely) no line in the stellar spectrum and the incident intensity is higher. Hence, the relative contribution of Raman-scattered light to the total reflected light is enhanced at wavelengths corresponding to prominent absorption lines, thus filling-in these lines (i.e. reducing their depth) in the reflected light. 

This additional component in the reflected light at the wavelengths of absorption lines in the stellar spectrum, compared to the surrounding continuum, causes positive peaks in the albedo spectrum, which we call albedo enhancements. The effect is most pronounced for the strongest absorption lines in the incident spectrum.

\item \textbf{Raman ghosts}: The intensity of Raman-scattered light is reduced at wavelengths corresponding to Raman-shifted positions of strong absorption lines in the incident spectrum. Therefore, the total intensity of the reflected light at these wavelengths is reduced compared to the surrounding continuum, creating weak dips in the reflected spectrum, and consequently in the planet's albedo, called Raman ghosts.
\end{enumerate}

The Raman effect occurs only on scattered photons, hence its signatures can only be observed in the reflected light, and not in transmission spectra of exoplanets.

The importance of Raman scattering in planetary atmospheres was first discovered in the 1960s. \cite{Brinkmann1968} used it to explain the observed filling-in of solar Fraunhofer lines in the Earth's atmosphere, also known as the Ring effect \citep{GraingerRing1962}. Studies of Raman scattering in the context of atmospheres of other planets began soon thereafter \citep{Wallace1972,Belton1971,Belton1973,Price1977,CochranTrafton1978,Cochran1981a,Cochran1981b} and in the next few decades the effects of Raman scattering have been observed in all gas planets in the Solar System \citep[see e.g.][]{Karkoscka1994,Yelle1987, BetremieuxYelle1999, Courtin1999}.

To include the effects of Raman scattering in the atmospheric models, we follow the method presented in \cite{BetremieuxYelle1999}. They studied Raman scattering in the atmosphere of Jupiter, which they observed using the Hubble Space Telescope. By comparing the observations with models, they identified specific Raman features and used them to determine the rotational distribution of H$_2$ in the Jupiter's atmosphere and the H$_2$ ortho-para ratio. Their models were computed using the publicly available radiative transfer code DISORT \citep{Stamnes1988}, which they modified to include the effects of multiple Raman scattering by H$_2$. In this study, we also use DISORT, to which we have made similar modifications in order to include the treatment of Raman scattering.

For the purpose of studying the atmosphere of Neptune, \cite{Sromovsky2005} developed the first method for treating Raman scattering that includes polarization effects. This accurate method was then used to evaluate several approximate methods for including Raman scattering in radiative transfer calculations that were developed with the goal of reducing computational costs associated with rigorous treatment of Raman scattering. The approximation introduced by \cite{Pollack1986} is often used to treat Raman scattering in modeling of reflected light and albedos of exoplanets \citep[e.g. ][]{Marley1999,Sudarsky2000,Cahoy2010}. \cite{Marley1999} use the Pollack approximation with correction terms calculated for a 6000-K blackbody. The albedo spectra produced by their models do not show distinct Raman features that are the focus of our study because a blackbody spectrum does not contain any lines.

\section{Methods}
\label{sec:methods}

\subsection{Model Atmospheres}

\label{sec:tp_profiles}

We investigate the signatures of Raman scattering imprinted on the geometric albedo of H$_2$- and N$_2$-dominated atmospheres. We focus our attention on the near-ultraviolet and blue end of the visible spectrum (the wavelength range $3000 \leq \lambda \leq 5000$ \AA) because the Raman scattering cross section has a strong wavelength dependence ($\sigma \propto \lambda^{-4}$) and the effects that we investigate become weaker as the wavelength increases. Compared to longer wavelengths, this part of the spectrum is relatively clean of absorption lines from atoms and molecules present in the atmosphere. The most notable absorption lines are those of sodium and potassium at $\sim 330$~nm and $\sim 404$~nm, respectively. The abundances of Na and K atoms in monatomic form strongly depend on the presence and structure of clouds in the atmosphere. We simplify our model atmospheres by neglecting absorption and including only opacity due to (Rayleigh and Raman) scattering. This allows us to more easily identify and study the effects of Raman scattering alone, which is the main focus of this paper.

For our main model atmosphere we assume that it is composed entirely of molecular hydrogen $^1$H$_2$ (90\% by number) and helium $^4$He (10\%). Both hydrogen and helium contribute to Rayleigh scattering, but only hydrogen molecules Raman scatter in the wavelength range that we are interested in. The mean molecular weight of this atmosphere is set to $\mu=2.3$~m$_p$ and the surface gravity is $g=10^3$~cm~s$^{-2}$. 

We use three different temperature-pressure (T-P) profiles for clear (cloud-free) H$_2$/He atmospheres: 
\begin{enumerate}
\item Hot atmosphere: T-P profile of the cloud-free case from \cite{Heng2012}. The temperature ranges from 1350~K to 1600~K. 

\item Warm atmosphere: T-P profile of the cloud-free model for warm Neptunes from \cite{Cahoy2010} (i.e. the Neptune-like planet at 0.8 AU from the host star) with temperature in the range $\sim 300 -1000$~K. 
\item Cold atmosphere: T-P profile of Neptune from \citet[Fig. 3]{Sromovsky2005} with temperature in the range $\sim 50 - 300$~K.
\end{enumerate}
The cloud-free atmospheres have a high-pressure (100 bar) bottom boundary with zero albedo. 

To assess the influence of thick clouds on the visibility of Raman features, we approximate the presence of a cloud deck at a given pressure by terminating the atmosphere and placing a Lambertian bottom boundary of our model atmosphere at that pressure level. Clouds can have a range of reflective/absorbing properties, which we incorporate by assigning different albedo values to the bottom boundary.

Additionally, we make a model atmosphere composed entirely of molecular nitrogen $^{14}$N$_2$, with a bottom boundary with albedo equal to 0.3 at a pressure level of 1-3 bar, and a TP profile matching that of the Earth's atmosphere\footnote{We use the data from the Planetary Atmospheres Node \url{http://atmos.nmsu.edu/planetary_datasets/earth_temppres.html}.}. Even though nitrogen-dominated atmospheres will most likely have other constituents that contribute to Rayleigh and Raman scattering (for example O$_2$ in the case of the Earth's atmosphere), we include only N$_2$ in order to investigate the effects of Raman scattering from this molecule alone. This analysis serves primarily as a demonstration that Raman scattering can be used to find signatures of gases other than H$_2$. Modeling realistic atmospheres of smaller planets is beyond the scope of this paper. The mean molecular weight for the N$_2$ atmosphere is $28.0$ m$_p$, and the surface gravity of the planet is $g=10^3$~cm~s$^{-2}$.

\subsection{Radiative Transfer Calculations}

We perform radiative transfer calculations using a discrete ordinate algorithm DISORT\footnote{\url{ftp://climate1.gsfc.nasa.gov/wiscombe/Multiple_Scatt/}} \citep{Stamnes1988}. The code was designed to perform monochromatic unpolarized radiative transfer in a scattering, absorbing and emitting plane-parallel medium. The medium is divided into $N_{lyr}$ homogeneous layers, separated by $N_{lyr}+1$ levels. The user defines the optical depth ($\tau$) and the single-scattering albedo ($\omega$) for each layer. The medium can be illuminated by a parallel beam of radiation at the top boundary, at an angle of incidence specified by the user (polar angle $\Theta_{\mathrm{in}}$, azimuth angle $\Phi_{\mathrm{in}} =0$). The code computes radiation intensities at user-selected angles and levels. All our calculations are performed with 50 computational levels ($N_{lyr}=49$) and 16 streams. 

To define computational levels for each model atmosphere used in the radiative transfer calculations, we start by choosing the minimum and maximum pressure. For minimum we choose $P_{\mathrm{min}}=10^{-4}$ bar. We tried a few different values of $P_{\mathrm{min}}$ (in the range $10^{-5} - 0.01$ bar) and the final output of the calculation did not change significantly. As mentioned before, the maximum pressure is $P_{\mathrm{max}}=100$ bar for the clear atmospheres, 0.1-1 bar for the cloudy H$_2$/He, and 1-3 bar for the nitrogen-dominated atmospheres. 

We divide the specified pressure range into computational pressure levels evenly distributed in logarithmic space. The temperature of each level is calculated using the temperature-pressure profile of the atmosphere. For each layer we compute its optical depth as
\begin{equation}
\tau_i = \frac{\sigma(T_i)}{\mu g}\Delta P_i \ \mbox{,}
\end{equation}
where is $\mu$ the mean molecular mass, $g$ is the surface gravity of the planet, and $\sigma$ is the total scattering cross section (with contributions from Rayleigh and Raman scattering) calculated for the local temperature of that layer. The temperature of a layer is calculated as the average of the corresponding upper- and lower-level temperature, and the change in pressure across a layer $\Delta P_i$ is just the difference in pressure between the upper and the lower level.

Single-scattering albedo is computed for every computational layer and level using cross sections for Rayleigh and Raman scattering by H$_2$ and Rayleigh scattering by He atoms:
\begin{equation}
\omega(\lambda) = \frac{f_\mathrm{H_2}\sigma_\mathrm{H_2,Ray}(\lambda)+f_\mathrm{He}\sigma_\mathrm{He,Ray}(\lambda)}{f_\mathrm{H_2}\sigma_\mathrm{H_2,Ray}(\lambda)+f_\mathrm{He}\sigma_\mathrm{He,Ray}(\lambda)+f_\mathrm{H_2}\sigma_\mathrm{H_2,Ram}(\lambda)}\ \mbox{,}
\end{equation}
where $f$ denotes the number fraction of each species. In a pure N$_2$ atmosphere we use only $\sigma_\mathrm{N_2,Ray}$ and $\sigma_\mathrm{N_2,Ram}$. We take into account rotational and (ro-)vibrational Raman transitions from the ground electronic and ground vibrational level of the molecule. The initial rotational levels range from $J=0$ to $J=9$ for hydrogen, and from $J=0$ to $J=25$ for nitrogen. They are populated according to the Boltzmann distribution at the local temperature of the atmosphere $T$:
\begin{equation}
f_J = \frac{g_J}{Z}e^{-E_J/k_BT} \ \mbox{,}
\label{eq:boltzmann}
\end{equation}
where $f_J$ is the fraction of molecules that are in a state with the rotational quantum number $J$,  $Z$ is the partition sum, $E_J$ is the energy of the initial state and $g_J$ its statistical weight given by
\begin{eqnarray}
g_J (\mathrm{H_2}) &=& \begin{cases}
2J+1 &\text{for $J$ even}\\
3(2J+1) &\text{for $J$ odd} \ \mbox{,}
\end{cases}\\
g_J (\mathrm{N_2}) &=& \begin{cases}
6(2J+1) &\text{for $J$ even}\\
3(2J+1) &\text{for $J$ odd} \ \mbox{.}
\end{cases}
\end{eqnarray}
The final molecular states are either in the ground or the first excited vibrational level. For more details on the cross sections used, see Appendix \ref{sec:cross_sections}.

The publicly available version of the DISORT code (we use version 1.3 written in Fortran) does not include treatment of Raman scattering. We modify the code by including an additional source term due to Raman scattering ($S_\mathrm{Ram}$) into the radiative transfer equation:
\begin{eqnarray}
\mu \frac{dI_\lambda(\lambda, \hat{k})}{d\tau}=& &I_\lambda(\lambda, \hat{k}) - S_\mathrm{Ram}(\lambda) \\
\nonumber &-& \frac{\omega(\lambda)}{4\pi}\oint d\Omega P(\hat{k}, \hat{k}^\prime)I_\lambda(\lambda, \hat{k}^\prime) \ \mbox{,}
\end{eqnarray}
where $I_\lambda(\lambda, \hat{k})$ is the specific intensity of radiation per unit wavelength, at wavelength $\lambda$, propagating in the direction of $\hat{k}$, $\mu$ denotes the cosine of the emission angle, $\tau$ is the optical depth, and $P(\hat{k}, \hat{k}^\prime)$ is the (Rayleigh) scattering phase function.

The only source term included in the standard version of DISORT describes thermal emission for each layer of the atmosphere. Because we are focusing on the wavelength range in which thermal emission is not very important ($\lambda \lesssim 5000$ \AA), we can modify this term to represent the Raman emission instead. As a result, our Raman source function is isotropic since it comes as a substitute for thermal emission, which was constructed with that property. The Raman emission is proportional to the mean radiation intensity at the Raman-shifted wavelength $J_\lambda(\lambda_j^\prime)$, the Raman cross section for the transition in question $\sigma_\mathrm{Ram}(\lambda_j^\prime)$ and the fractional population of the initial molecular state $f_j$:
\begin{equation}
S_\mathrm{Ram} = \sum_j J_\lambda(\lambda_j^\prime) \frac{f_j\sigma_\mathrm{Ram}(\lambda_j^\prime)}{\sigma_\mathrm{tot}(\lambda)}\left(\frac{\lambda_j^\prime}{\lambda} \right)^3 \ \mbox{,}
\label{eq:ram_source}
\end{equation}
The sum is taken over all relevant Raman transitions. The wavelength of the outgoing photon ($\lambda$) is related to the incident wavelength ($\lambda_j^\prime$) through 
\begin{equation}
\frac{1}{\lambda} = \frac{1}{\lambda_j^\prime}-\Delta\nu_j  \ \mbox{,}
\end{equation}
where $\Delta\nu_j$ denotes the Raman shift in wavenumber (i.e. inverse wavelength) for the $j$-th transition. The factor $(\lambda_j^\prime/\lambda)^3$ in the source term originates from the fact that Raman scattering conserves the number of photons, but redistributes them in wavelength space. In the radiative transfer equation we define the intensity in terms of energy per unit wavelength, hence we need to take that redistribution into account. One factor of $\lambda_j^\prime/\lambda$ comes from the change in the energy of the photons due to scattering and another $(\lambda_j^\prime/\lambda)^2$ comes from the change in the wavelength range over which the photons get redistributed over (note that the Raman shift is constant in frequency, but changes in wavelength-space).

To calculate the Raman source term in each layer of the atmosphere at a given wavelength, we need to know the mean radiation intensity in that layer at Raman-shifted wavelengths, for all relevant Raman transitions. Hence, we need to cover the entire spectral range wavelength by wavelength, record the mean intensities and then use them to compute Raman intensities at different wavelengths. We start this procedure at the shortest wavelength (3000 \AA) and run the modified DISORT code without any Raman source terms. At each layer we record the mean intensity. After that, we run the code for longer wavelengths, up to 5000 \AA\ with 1 \AA\ spacing, now including the emission due to Raman scattering at shorter wavelengths. This accounts for Raman scattering transitions that lower the photon energy (i.e. the Stokes transitions). After we complete calculations over the entire wavelength range, we repeat them once again, this time having at our disposal the mean intensities over the entire wavelength range. This allows us to include the anti-Stokes Raman transitions (i.e. the Raman shifts to shorter wavelengths) as well.

The main output of the code is the radiation intensity at the top boundary in a specified direction (angle). To obtain the disk-integrated intensity we compute the outgoing intensity at several angles distributed across the planetary disk, as described in the next section and the Appendix \ref{sec:quadrature}.

\subsection{Geometric Albedo Calculation}
\label{sec:albedo_calc}

Geometric albedo $A_g$ is a measure of the reflectivity of a planet. It is defined as the ratio of the integral brightness of the planet at full phase to the brightness of a perfect Lambert disk of the same size and at the same distance from the star \citep[see e.g.][]{hapke2012}. Geometric albedo can be calculated as the ratio of the total emergent intensity of the planet at zero phase angle $j(0)$ and the incident stellar flux $\pi I_\mathrm{inc}$ at the planet-star distance \citep{Sudarsky2000,MadhusudhanBurrows2012}:
\begin{equation}
A_g = \frac{j(0)}{\pi I_\mathrm{inc}} \ \mbox{.}
\end{equation}
Geometric albedo can be used to estimate the planet-star flux ratio observed on Earth as a function of the phase angle $\alpha$ (the angle between the line connecting the star and the planet and the line connecting the planet and the observer)
\begin{equation}
\frac{F_p}{F_*} = A_g \left(\frac{R_\mathrm{pl}}{a}\right)^2 \Phi(\alpha)\ \mbox{,}
\label{eq:pl_st_flratio}
\end{equation}
where $R_\mathrm{pl}$ is the radius of the planet, $a$ is the planet-star distance and $\Phi(\alpha)=j(\alpha)/j(0)$ is the phase function normalized at full phase.

In order to compute the total intensity of the planet $j(\alpha)$ at the phase angle $\alpha$, we need to integrate the emergent specific intensity over the entire illuminated surface of the disk. To that end, we follow the procedure described in \cite{Horak1950, HorakLittle1965, MadhusudhanBurrows2012}. In the following equations $I(\eta, \zeta)$ denotes the specific intensity emerging from the part of the planetary surface with coordinates (co-latitude, longitude)=$(\eta, \zeta)$. Disk-integrated intensity is then given by
\begin{equation}
j(\alpha) = \int_0^{\pi} d\eta \sin^2\eta \int_{\alpha-\pi/2}^{\pi} d\zeta I(\eta, \zeta) \cos\zeta \ \mbox{.}
\label{eq:disk_int1}
\end{equation} 
The cosine of the angle of incidence and reflection is $\mu_0 = \sin{(\eta)}\cos{(\zeta-\alpha)}$ and $\mu=\sin{(\eta)} \cos{(\zeta)}$, respectively. These two angles are equal at full phase ($\alpha=0$). To evaluate the integral in equation (\ref{eq:disk_int1}) numerically, first we make the following coordinate transformation:
\begin{eqnarray}
\psi &=& \cos\eta \\
\nu &=& \sin\zeta \\
\xi &=& \frac{2}{\cos{\alpha}+1}\nu + \frac{\cos{\alpha}-1}{\cos{\alpha}+1}
\end{eqnarray} 
to get
\begin{equation}
j(\alpha) = \frac{\cos{\alpha}+1}{2}\int_{-1}^{1} d\psi \sqrt{1-\psi^2} \int_{-1}^{1} d\xi I(\psi, \xi) \ \mbox{.}
\end{equation}
This double integral can be evaluated using standard quadrature methods \citep{HorakLittle1965}
\begin{equation}
j(\alpha) = \frac{\cos{\alpha}+1}{2} \sum_{i=1}^n \sum_{j=1}^n a_i b_j I(\psi_i, \xi_j) \ \mbox{,}
\label{eq:quad}
\end{equation}
where $a_i$, $\psi_i$ are the Chebyshev weights and divisions, and $b_j$, $\xi_j$ are the Gaussian weights and divisions. We use quadrature of order $n=6$ with coefficients that are listed in Appendix \ref{sec:quadrature}. 

Because of the symmetry of the $\alpha=0$ problem, for $n=6$ we only need to evaluate $I(\psi_i, \xi_j)$ at nine different points on the planet's disk. Hence, we run our radiative transfer code for nine different angles of incidence of the radiation beam, and record the emergent specific intensity at the same angle\footnote{Due to conventions used to define angles specifying the direction of radiation propagation, DISORT variables PHI0 and PHI which designate the azimuthal angles of the incident beam and the output, need to have a 180$^\circ$ offset in order to be oriented in the same direction.}. We sum up the results according to equation (\ref{eq:quad}) to get the disk-integrated emergent intensity. We divide that with the incident irradiance to get the geometric albedo $A_g$.

\section{Results}
\label{sec:results}

In order to test our implementation of Raman scattering in DISORT, we perform two tests to compare our results to those reported by \cite{BetremieuxYelle1999} and \cite{Sromovsky2005}. The results of these relatively simple test cases also nicely illustrate how Raman scattering affects the reflected light from a planet. Firstly, we look at the Raman-scattered light produced by a monochromatic source and then we illuminate the atmosphere with a flat photon spectrum with a single absorption line. After these simple test cases, we calculate the geometric albedo resulting from irradiating different model atmospheres with the solar spectrum \citep{Kurucz2005}\footnote{\url{http://kurucz.harvard.edu/sun/irradiance2005/irradthuwl.br_2nmresamp}}.

\subsection{Test Case I: Monochromatic Source}

\begin{figure}
\centering
\includegraphics[width=0.5\textwidth]{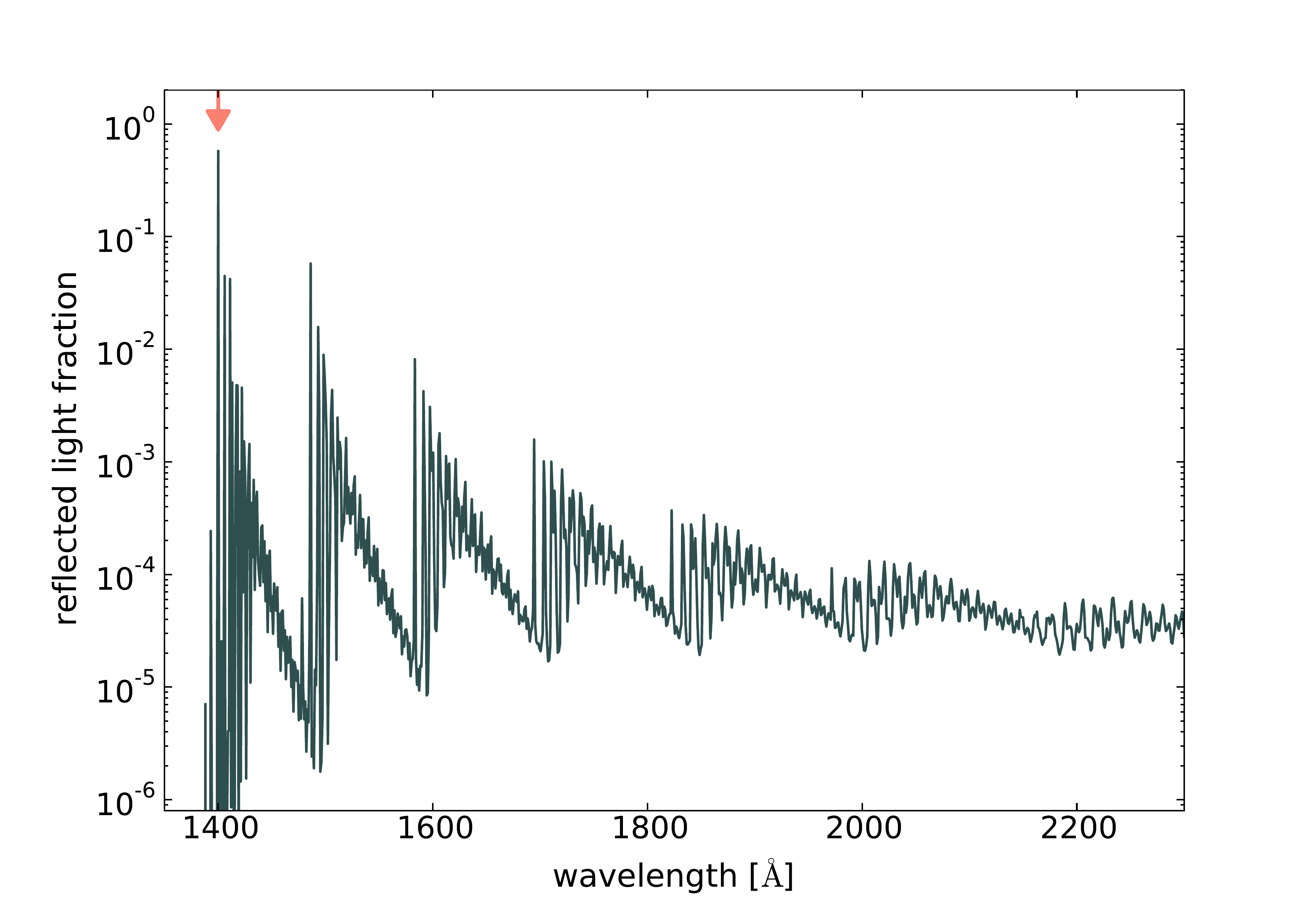}
\caption{Reflected light for the case of monochromatic ($\lambda=1400$ \AA) radiation illuminating a clear 10 bar-deep H$_2$/He atmosphere. The arrow points at the largest peak which is the reflected light due to Rayleigh scattering at $\lambda=1400$~\AA. It accounts for about 60\% of the total reflected light. Everything else comes from Raman scattering. The peaks around the incident wavelength are due to pure rotational Raman transitions. Peaks at wavelengths longer than $\sim$1480~\AA\ are due to various combinations of vibrational and rotational Raman transitions.}
\label{fig:monotest}
\end{figure}

In this case, a H$_2$/He atmosphere is illuminated from the top with a beam of monochromatic light of wavelength $\lambda_\mathrm{beam}=1400$~\AA. There is no incident radiation at other wavelengths. The bottom boundary is at the pressure of 10 bar and its albedo is set to zero. We record the total flux emerging from the top of the atmosphere. Fig.~\ref{fig:monotest} shows the emerging flux as a function of wavelength, normalized to the incident beam. The largest peak (at 1400~\AA), accounting for $\sim 60$\% of the reflected flux, corresponds to Rayleigh-scattered light, everything else is reflected light due to Raman scattering. The first group of peaks represents photons that are produced by pure rotational Raman transitions -- anti-Stokes for the peaks at shorter wavelengths than the incident radiation and Stokes transitions for the peaks at longer wavelengths. The second group of peaks, starting around 1480~\AA, and all consequent series correspond to combinations of vibrational and rotational transitions. 

\cite{BetremieuxYelle1999} and \cite{Sromovsky2005} perform similar tests of Raman scattering of monochromatic radiation. Our results shown in Fig. \ref{fig:monotest} look very similar to results from both these papers, giving us confidence that our treatment of Raman scattering is sound. We do not expect our result to be in perfect agreement with theirs due to differences in the assumed atmospheric properties and taking into account Raman transitions from a different number of initial molecular states (we allow more excited rotational levels of the molecule to be initially populated, not just levels with $J=0$ and $J=1$).

\subsection{Test Case II: Flat Spectrum with an Absorption Line}

\begin{figure*}
\centering
\includegraphics[height=0.27\textheight]{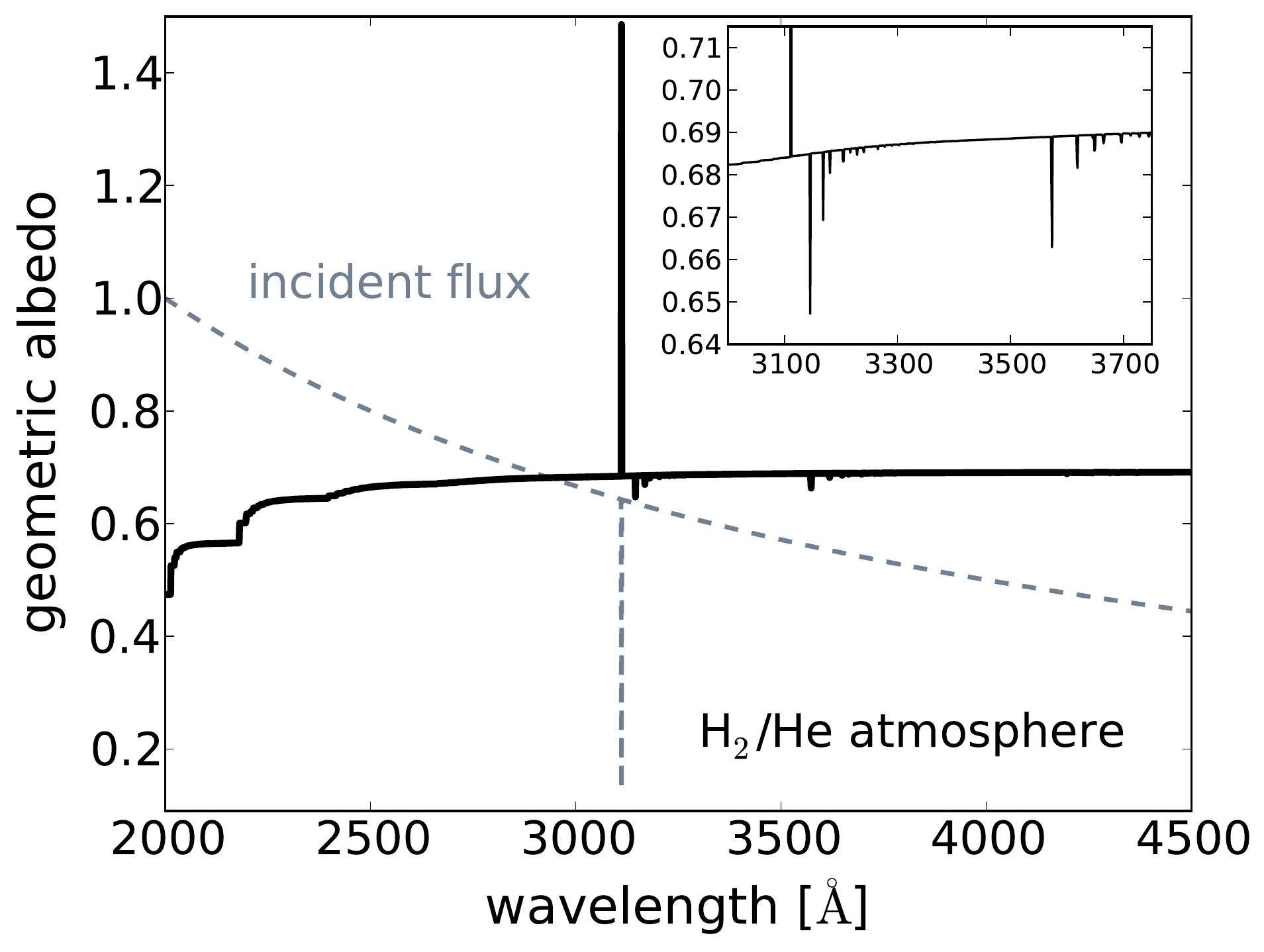}
\includegraphics[height=0.27\textheight]{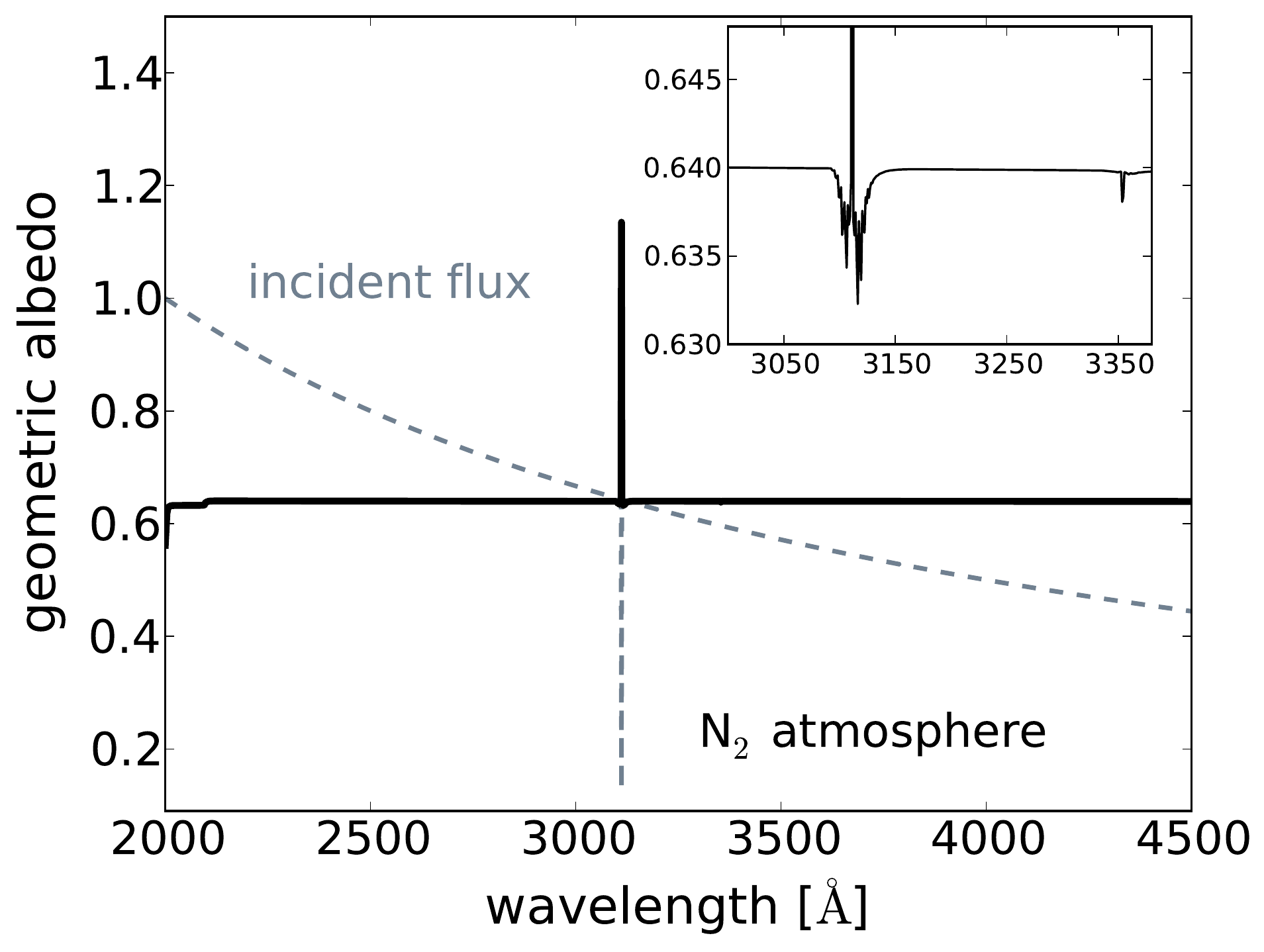}
\caption{Left panel: Solid line shows the geometric albedo for a clear, cold H$_2$/He atmosphere illuminated by a radiation beam with a flat photon spectrum with an absorption line at $\lambda \sim 3100$ \AA\ (gray dashed line). The albedo is strongly enhanced at the wavelength of the absorption line. Raman ghost lines appear at Raman-shifted wavelengths (the inset shows a zoomed-in view of the ghost lines). Right panel shows the same for a clear 100-bar atmosphere of pure N$_2$. The structure and positions of Raman ghost lines are different for the two molecules, H$_2$ and N$_2$, reflecting the structure of energy levels for each molecule. Hence, if observed, Raman ghost lines could be used to spectroscopically identify the composition of the scattering medium.}
\label{fig:abstest}
\end{figure*}

The second test case matches the one presented in \cite{Sromovsky2005}. A 100-bar deep, cold-model (as defined in Section \ref{sec:tp_profiles}) atmosphere is illuminated by a beam of radiation with an absorption line at $\lambda = 3100$ \AA\ inserted into an otherwise flat spectrum (constant $\lambda I_{\lambda}$, with $I_{\lambda}$ normalized at 2000 \AA). Fig.~\ref{fig:abstest} shows the incident spectrum (dashed line) and the geometric albedo spectrum (solid line) calculated by our radiative transfer code. Our result for the hydrogen-dominated atmosphere looks very similar to that of Sromovsky, although we cannot reproduce the exact values of the albedo because we do not include the effects of radiation polarization.

This test case demonstrates the main features of Raman scattering that were discussed in Section \ref{sec:raman}. At the position of the absorption line in the incident spectrum, there is a strong peak in the geometric albedo spectrum caused by photons that are Raman shifted into that wavelength. Dips in the geometric albedo -- Raman ghosts of the absorption line -- are less prominent, yet noticeable. Fig. \ref{fig:abstest} shows that the positions of ghost lines in an H$_2$-dominated atmosphere are different from those in the atmosphere of N$_2$. Raman shifts are determined by the rotational and vibrational energy levels of the molecule and hence the relative shifts of ghost lines are unique for every species. This demonstrates how observations of Raman ghost lines could be used to spectroscopically identify homonuclear diatomics like H$_2$ and N$_2$ that otherwise do not have prominent spectral features in this wavelength range.

Intensities of individual ghost lines depend on the cross section for that particular transition and on the population of the initial state, governed by the Boltzmann distribution. By measuring the relative intensities of two ghost lines produced by the same feature in the incident spectrum, the temperature of the medium could be inferred.

\subsection{Hydrogen Atmospheres Irradiated by Solar Spectrum}
\label{sec:albedos_H2}

\begin{figure}
\centering
\includegraphics[width=0.5\textwidth]{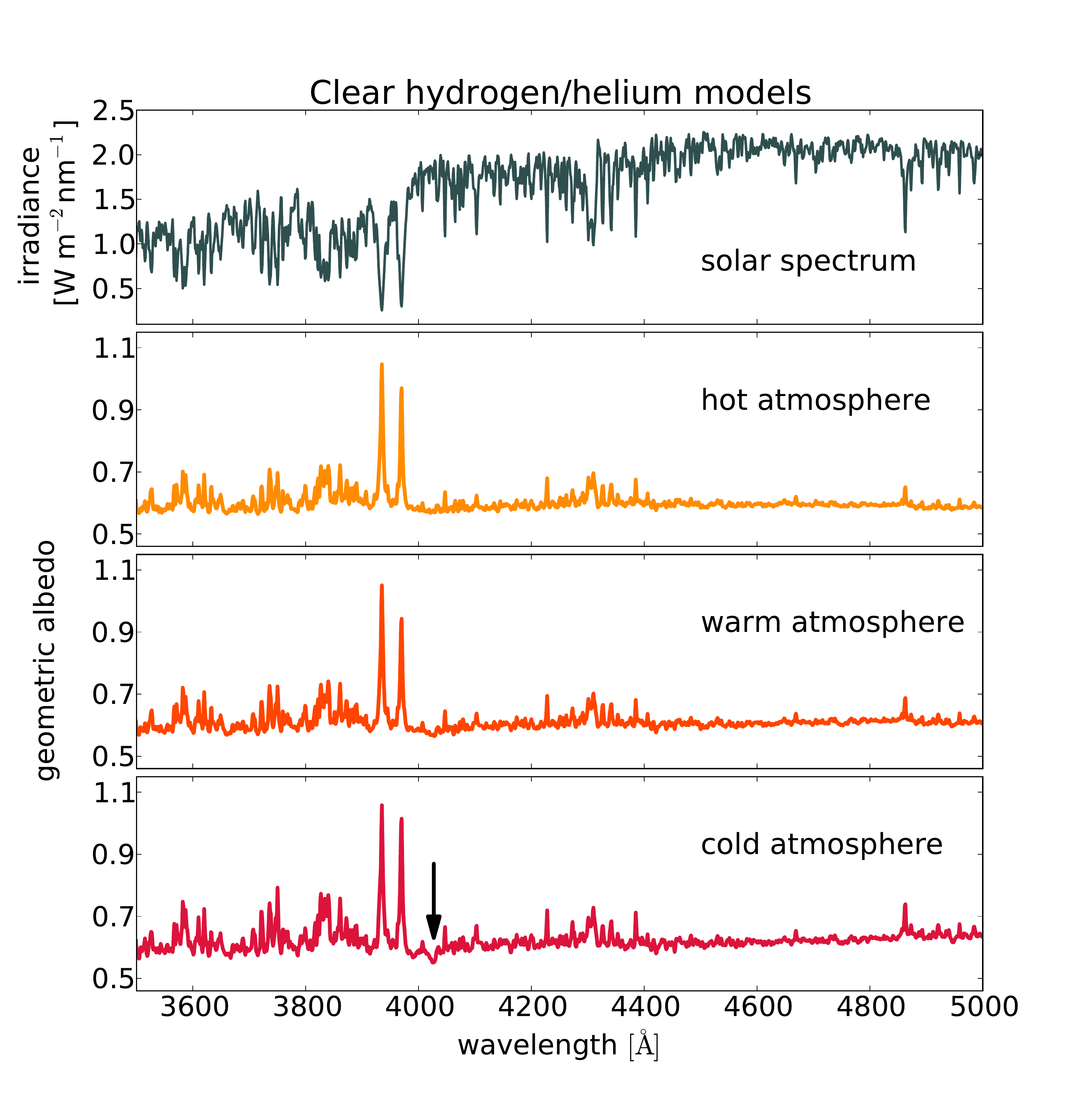}
\caption{Three lower panels show the geometric albedo spectra calculated for clear 100-bar deep H$_2$/He atmospheres at different temperatures. The incident radiation has a solar-type spectrum (top panel). In an atmosphere with only Rayleigh scattering, the albedo spectrum would be flat, hence all the features present in these spectra are due to Raman scattering. The most prominent Raman features are the albedo enhancements at wavelengths corresponding to strong absorption lines in the solar spectrum. The intensity of individual Raman ghost lines, such as the feature indicated by an arrow in the bottom panel, is much weaker and it is more significantly affected by the temperature of the atmosphere.}
\label{fig:albedo_temperature}
\end{figure}

In Fig.~\ref{fig:albedo_temperature} we show the incident spectrum and the computed albedo spectra for three models of clear (cloud-free) H$_2$/He atmospheres. The three models have different temperature-pressure profiles, as described in Section~\ref{sec:tp_profiles}. The albedo spectra show enhancements of the order of tens of percent at wavelengths at which the incident spectrum has absorption lines. In the cores of the most prominent lines, the Fraunhofer H and K lines of ionized calcium at 3969 and 3934~\AA, the geometric albedo can be almost twice as high compared to the surrounding continuum.

The albedo spectra look very similar for all three cloud-free models, although there are some differences due to the temperature dependence of the population of molecular rotational levels. Raman transitions originating from different initial levels have (slightly) different cross sections, changing the overall strength of the Raman effect with temperature. The cross sections are highest for low-energy initial states (for example, see Fig.~\ref{fig:cross_sections} -- the cross section for the rotational S-branch transition from level $J=0$ is higher than the cross section for the initial level  $J=1$), making the Raman effect somewhat stronger in colder atmospheres.

The position of ghost lines also depends on the initial state of the molecule, and hence the temperature of the medium. The most prominent ghost feature in the hydrogen spectrum is located at $\lambda \approx 4027$~\AA\ (indicated by an arrow in the bottom panel of Fig. \ref{fig:albedo_temperature}). Two strong ghost lines nearly overlap at this wavelength -- a ghost originating from an $S(0)$ Raman transition of the calcium H line, and the $S(1)$ transition from the calcium K line.  In our cold atmosphere model, almost all hydrogen is in the $J=0$ and $J=1$ levels, maximizing the intensity of the ghost lines originating from these two states. In this model, the ghost feature at $\sim$4027~\AA\ causes a $\sim 10$\% decrease in the albedo. At higher temperatures, the initial distribution of molecular states is broader, making individual ghost lines originating from low-energy states less intense.

\begin{figure*}
\centering
\includegraphics[height=0.33\textheight]{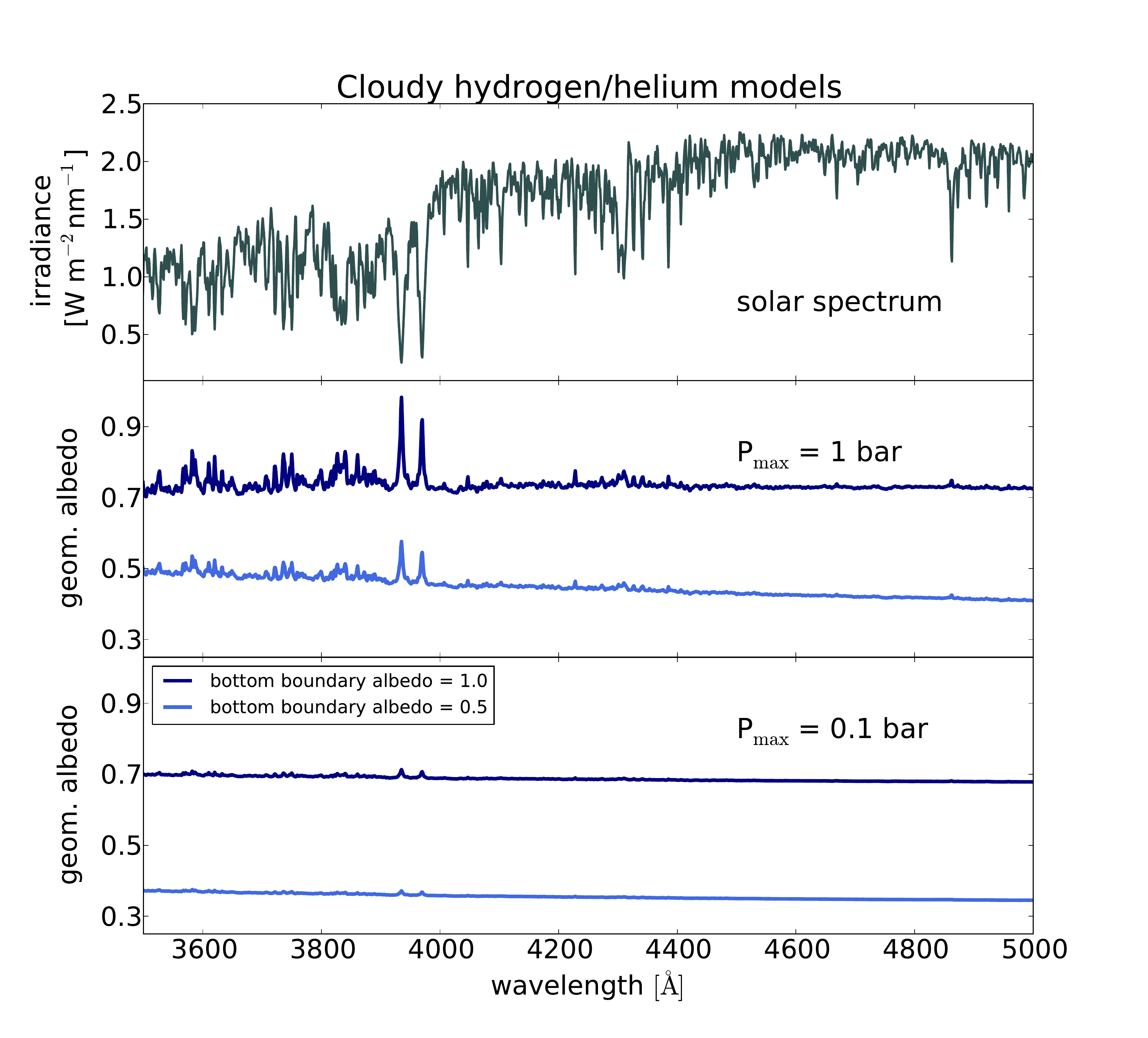}
\includegraphics[height=0.33\textheight]{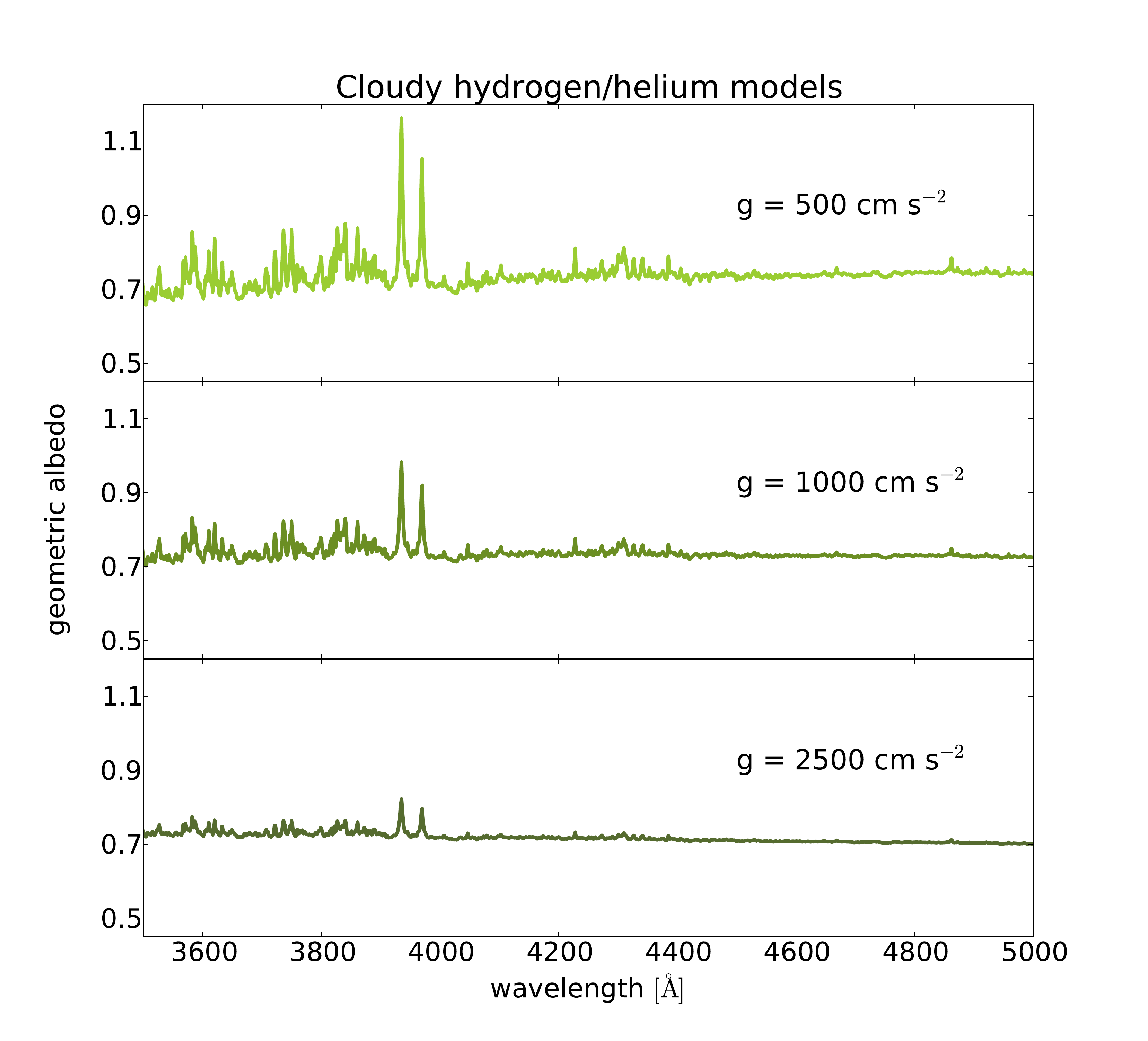}
\caption{\textit{Left:} Middle and bottom panels show geometric albedos for H$_2$/He atmospheres with a perfectly reflecting (albedo = 1.0) and partially absorbing (albedo = 0.5) bottom boundaries at the pressure value $P_\mathrm{max}$, representing a thick cloud deck. Spectral features due to Raman scattering are less prominent in shallow atmospheres because placing the cloud deck at higher altitudes (lower pressures) decreases the column of gas through which the radiation passes, thereby reducing the number of (Raman) scattering events.The column mass depends on the ratio of pressure and surface gravity, hence there is a degeneracy between these two quantities. The surface gravity of all these models is $10^3$~cm~s$^{-2}$. Top panel shows the incident spectrum. \textit{Right:} Geometric albedo for H$_2$/He atmospheres with the bottom boundary at $P_\mathrm{max}=1$~bar with albedo of 1.0, for planets with different values of surface gravity. }
\label{fig:albedo_pressure}
\end{figure*}

Placing a thick cloud deck at different altitudes (i.e. different values of pressure) affects the overall albedo spectrum even more than changing the temperature, as shown in Fig.~\ref{fig:albedo_pressure}. If a cloud deck is high in the atmosphere, the light passes through a smaller column of gas and experiences less scattering than when it travels through a clear atmosphere or an atmosphere with low-altitude clouds. Because the total column mass of gas above a certain level of the atmosphere is proportional to the ratio of the pressure at that level and the planet surface gravity, our results are degenerate with respect to these two quantities. Hence, in order to measure the pressure level of clouds using the intensity of Raman peaks, the surface gravity of the planet needs to be known. We demonstrate this in Fig.~\ref{fig:albedo_pressure} where we show that the intensity of Raman features can be altered by changing the pressure level of the bottom boundary, while keeping the surface gravity fixed (left panel), or by changing the surface gravity while the bottom-boundary pressure remains the same (right panel).

What affects the strength of Raman features is the number of molecules available to photons for scattering above the cloud deck -- having fewer Raman scattering events leads to an albedo spectrum in which Raman features are suppressed and the spectrum is dominated by Rayleigh scattering and the reflective properties of the clouds. Therefore, the strength of the Raman features can be used to infer the column mass of the scattering gas above the could deck and to put constrains on the presence and the altitude of clouds. In this optically thin regime, the albedo spectrum is much more sensitive to the properties of the planet than in the case of a clear, deep atmosphere where radiation can reach the pressure level at which the total optical depth (due to Rayleigh plus Raman scattering) becomes order unity. In the optically thin case,  the overall planetary albedo depends on the albedo of the bottom boundary (i.e. the cloud deck) -- more reflective bottom boundary results in a higher value of the geometric albedo continuum and larger Raman features as more photons get reflected from the bottom and have a second passage through the atmosphere on the way to the observer, allowing for more scattering evens to occur.

\subsection{Nitrogen Atmospheres Irradiated by Solar Spectrum}
\label{sec:albedos_N2}

\begin{figure}
\centering
\includegraphics[width=0.49\textwidth]{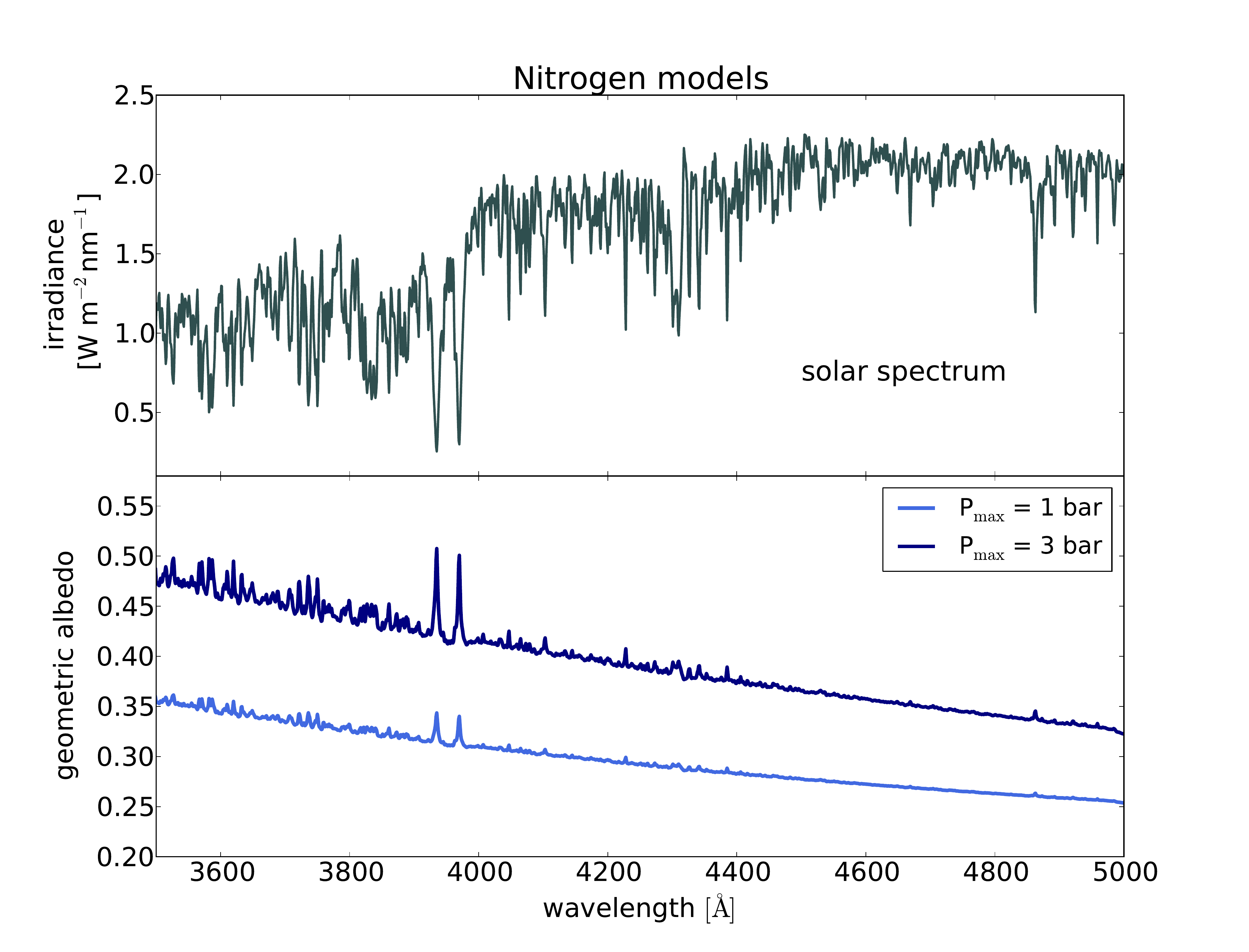}
\caption{Bottom panel shows the geometric albedo for the model atmosphere composed entirely of N$_2$. The bottom boundary with albedo 0.3 is placed at the pressure $P_\mathrm{max}$. Higher pressure indicates a deeper atmosphere, leading to more Raman scattering events and thus producing stronger Raman features in the albedo spectrum. Top panel shows the incident spectrum.}
\label{fig:albedo_nitrogen}
\end{figure}

We choose the parameters describing our nitrogen atmosphere so that they resemble the properties of Earth's atmosphere. We place the bottom boundary representing the surface of the planet at the pressure of 1~bar, and its albedo is set to 0.3. This choice of parameters leads to a relatively small column of the scattering gas, which is the reason why the albedo spectrum for this model shows only moderate effects of Raman scattering, as shown in Fig.~\ref{fig:albedo_nitrogen}. The effects are stronger for deeper atmospheres, as shown for the atmosphere ending at the pressure of 3~bar. In the case of a small planet with a shallow atmosphere, Raman features alone are unable to give information on whether the bottom boundary is an opaque cloud deck or the planet surface.

\section{Discussion}
\label{sec:discussion}

\subsection{Probing Exoplanet Atmospheres with Raman Scattering}

As shown in the previous section, the strength of Raman scattering features in the geometric albedo spectrum depends on the properties of the atmosphere. In this section, we discuss how observations of these features could be used to probe exoplanet atmospheres. We start from the most easily detectable Raman effects and move towards fainter features that require increasingly more precise observations of planetary albedo.

\subsubsection{Constraining the Pressure Level of the Cloud Top/Surface}

The most prominent and most easily detectable Raman features are the peaks in the geometric albedo at wavelengths corresponding to absorption lines in the incident spectrum. The albedo is enhanced at wavelengths corresponding to all lines, however the effect is most noticeable for strong lines, such as the Ca II H and K lines in the stellar spectrum. At these wavelengths the albedo can be enhanced by almost a factor of $\sim2$ in the case of a deep and clear H$_2$-rich atmosphere (Fig.~\ref{fig:albedo_temperature}) or by tens of percent in atmospheres with thick clouds at pressures around 1 bar (Fig.~\ref{fig:albedo_pressure}). In general, the strength of these features depends on the number of Raman scattering events, i.e. on the column density of the scattering molecule that the light goes through before being scattered into our line-of-sight. This could be used to probe the depth of the atmosphere and put constraints on the pressure level of clouds or the planetary surface, if the surface gravity of the planet is known. If the dominant source of opacity at short wavelengths is scattering on particles (aerosols), instead of molecules, Raman features are not expected to appear in the spectrum. 

\subsubsection{Identifying the Dominant Scatterer}

Raman scattering imprints `ghosts' of features present in the incident spectrum at shifted wavelengths in the reflected spectrum. By measuring the Raman shift of the lines, the scattering molecule can be identified. As shown in Fig.~\ref{fig:abstest}, the positions of Raman ghosts originating from the same line in the incident spectrum are different for H$_2$, N$_2$ and any other molecule. The intensity of individual ghost lines is small, on the order of a few percent of the albedo for the strongest lines. However, because Raman scattering occurs over the entire observed wavelength range, small dips in the geometric albedo should appear at frequencies shifted by a fixed amount with respect to \textit{all} stellar lines. Hence, it might be possible to get a statistical detection of the ghost lines and to measure the Raman shifts by doing a cross-correlation analysis of the incident spectrum and the geometric albedo in frequency-space \citep{Cochran1981a,Cochran1981b}. 

\begin{figure}
\centering
\includegraphics[width=0.5\textwidth]{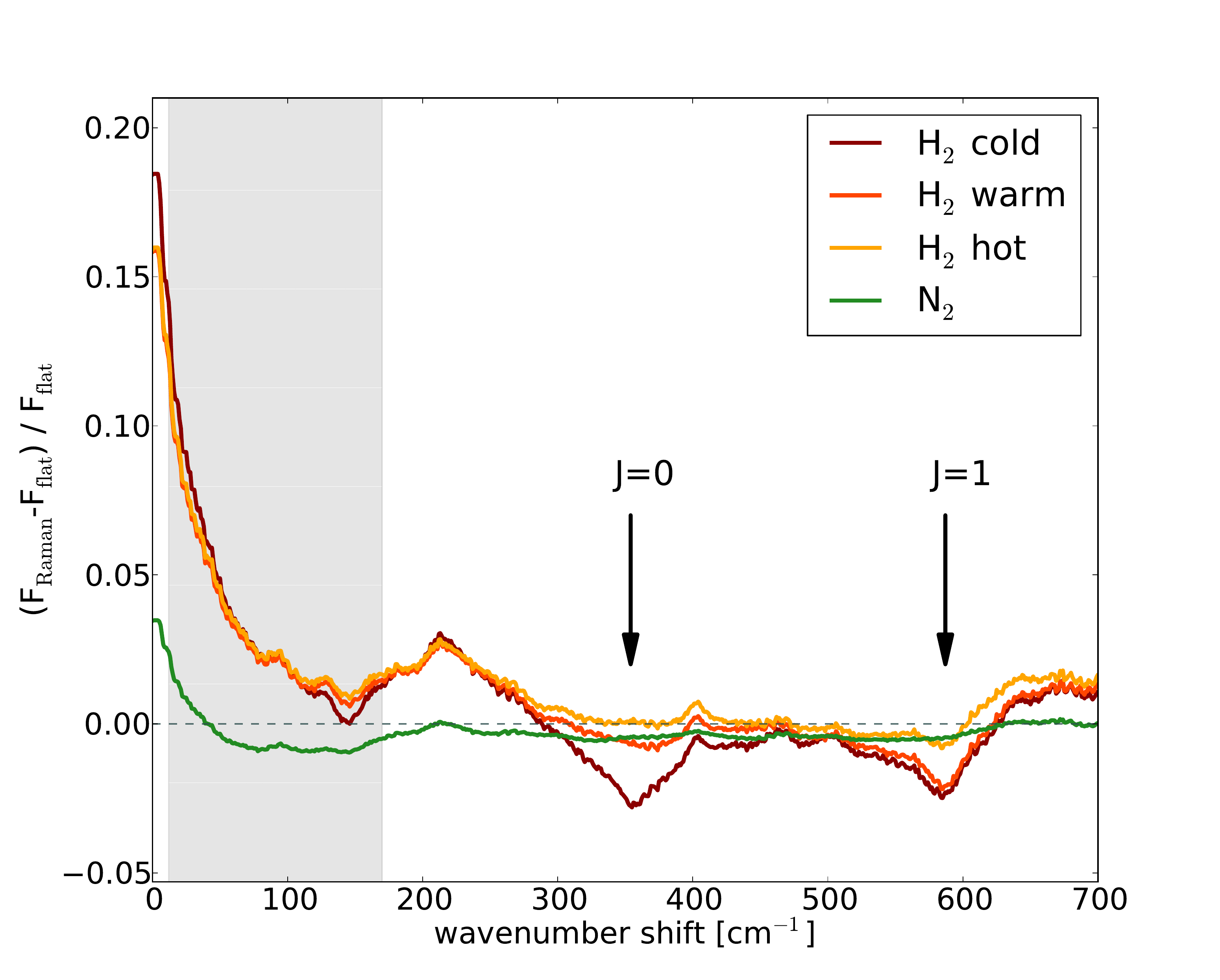}
\caption{Cumulative deviation from a flat albedo spectrum in the vicinity of prominent absorption lines in the incident spectrum. At very small wavenumber separations the albedo is enhanced due to the filling-in effect of Raman scattering. Dips in the albedo are Raman ghost lines. By detecting the position of Raman ghosts in the albedo spectrum, the identity of the scattering molecule can be inferred. Arrows mark the expected positions of the H$_2$ ghost lines due to rotational Raman transitions from initial states with $J=0$ and $J=1$. Hatched region marks the expected positions of the N$_2$ ghosts for $J=0,...,12$. The green curve shows a broad dip at these small wavenumber shifts and no discernible dips at the positions of H$_2$ ghost lines.}
\label{fig:identifying}
\end{figure}

In Fig.~\ref{fig:identifying}, we show such a cross-correlation with our simulated spectra. The $y$-axis shows the fractional difference between the planet spectrum with Raman features included ($F_\mathrm{Raman}$) and the planet spectrum with a flat albedo ($F_\mathrm{flat}$) showing no signs of Raman scattering. The $x$-axis shows the wavenumber ($1/\lambda$) offset from a strong line in the incident spectrum. We select $\sim 5$\% of the wavelength bins in the analyzed wavelength range that correspond to the strongest absorption lines in the solar spectrum. We analyze how the planet spectrum around each of these strong lines is affected by Raman scattering, sum up their relative contributions and present them in different colors for different model atmospheres. 

At very small wavenumber shifts (i.e. close to the cores of stellar absorption lines) the planetary flux is enhanced by Raman scattering by a few tens of percent -- the lines are `filled-in'. In hydrogen-dominated atmospheres we see two troughs in the planet flux due to Raman scattering, around $\sim 354$~cm$^{-1}$ and $\sim 587$~cm$^{-1}$. These flux decrements of the order of a few percent, correspond to Raman ghosts of strong stellar lines due to pure rotational Raman transitions from initial molecular states with $J=0$ and $J=1$. The troughs are strongest and most easily detectable for the cold atmosphere ($T\sim 100$~K) because these two molecular states account for the majority of H$_2$ molecules at such low temperatures. At higher temperatures more high-energy (high-$J$) states are populated and the intensity of individual ghosts is hence reduced. Raman shifts of molecular nitrogen are much smaller than those of hydrogen. Offsets corresponding to rotational Raman transitions from states with $J$ in the range 0-12 are shown in Fig.~\ref{fig:identifying} by the gray hatched region. Because the distribution of Raman shifts in N$_2$ is so broad, it is more difficult to discern. Another difficulty associated with small Raman shifts is that ghost lines can overlap with line wings, especially for strong, broad lines like the Fraunhofer H and K lines. Raman ghosts can blend with parent lines and cause merely a decrease in the filling-in effect, instead of showing up as discrete troughs in the planetary flux, as shown in the case of the N$_2$ model in Fig.~\ref{fig:identifying}.

\subsubsection{Constraining the Atmospheric Temperature}

Raman transitions from different initial states of the molecule have different Raman shifts, which is why an individual absorption line in the incident spectrum produces a whole forest of ghost lines in the reflected light, as shown in Fig.~\ref{fig:abstest}. The strength of a single ghost line depends on the total number of molecules in the corresponding initial state. For example, the intensity of a hydrogen ghost line resulting from a transition from the state with the rotational quantum number $J$ is
\begin{equation}
I(J) \propto N_\mathrm{H_2}(J) \ \mbox{.}
\end{equation}
There is a degeneracy between the overall abundance of H$_2$ molecules and the fractional population of individual rotational states:
\begin{equation}
N_\mathrm{H_2} (J) = N_\mathrm{H_2}^\mathrm{tot} f_J \ \mbox{.}
\end{equation}
The fractional population of states is governed by the Boltzmann distribution, given by equation (\ref{eq:boltzmann}). From measuring the strength of just one ghost line $I(J)$, we cannot say whether that line is strong because the overall abundance of the molecule ($N_{H_2}^{tot}$) is high or because the fractional population of the corresponding initial state ($f_J$) if high. However, if we could measure at least two ghost lines and take their ratio, the total number of molecules cancels out and what if left is the relative population of these two initial states, which is related to the temperature of the atmosphere in the following way:
\begin{equation}
\frac{I(J_1)}{I(J_2)} \propto e^{-(E_{J_1}-E_{J_2})/k_B T} \ \mbox{.}
\end{equation}
Because individual ghost lines are quite faint (up to a few percent of the albedo for the strongest lines) the uncertainties in the measured albedo should be well below a percent level to perform this type of measurement.

\subsection{Detection Feasibility}

In this section, we analyze the feasibility of observing Raman scattering in the atmospheres of nearby exoplanets, using the currently existing and the next generation observational facilities. First, we demonstrate that Raman features can be discerned in spectra of moderate spectral resolution.

\subsubsection{Raman Scattering Features at Different Spectral Resolutions}

\begin{figure}
\centering
\includegraphics[width=0.48\textwidth]{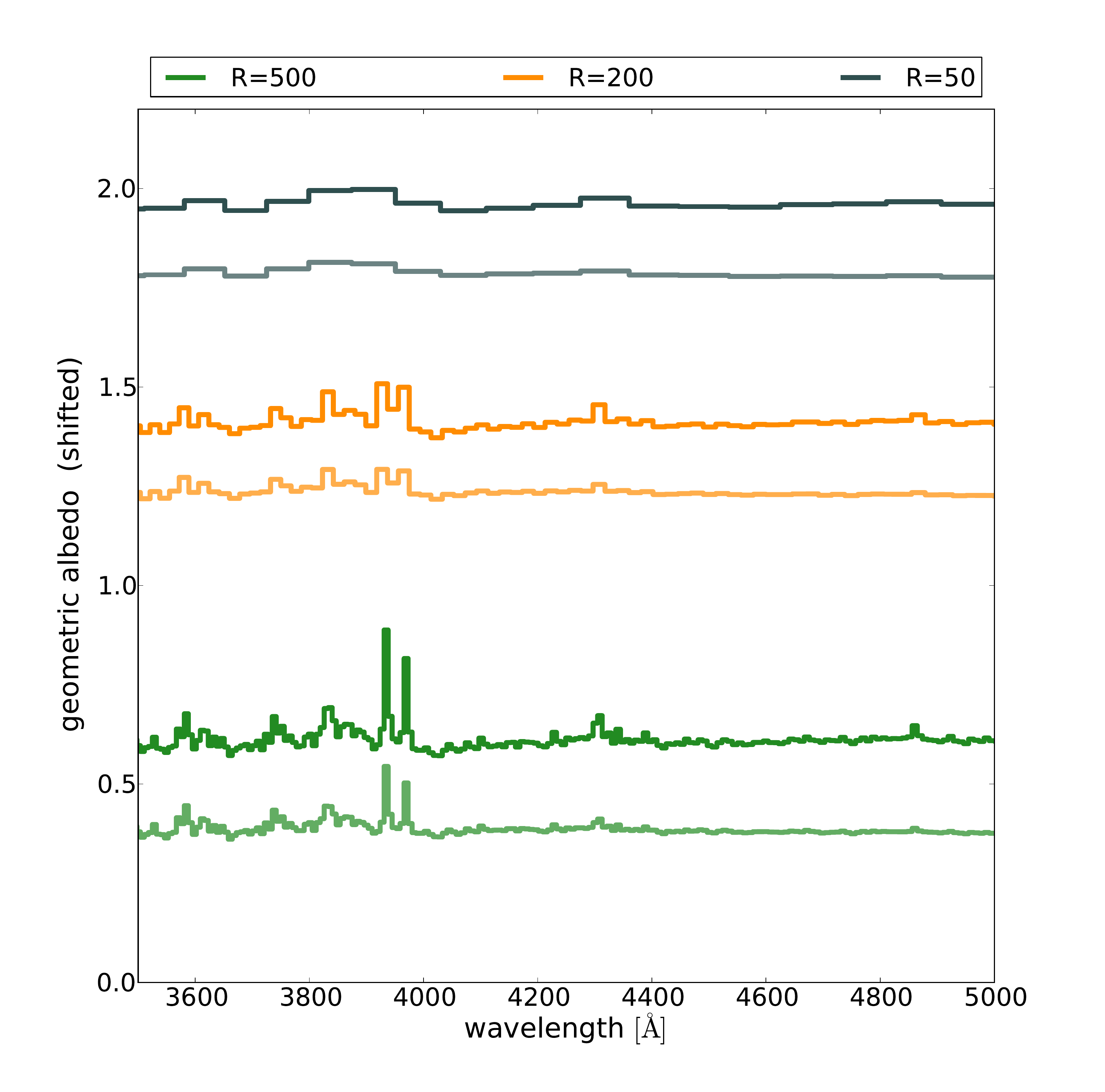}
\caption{Calculated albedo spectra degraded to lower spectral resolutions of $R=500$ (green), $R=200$ (yellow) and $R=50$ (gray). For every value of $R$, there is a pair of lines -- the lighter line shows the albedo for a cloudy ($P_{max}=1$ bar) H$_2$/He atmosphere and the darker line is for a clear atmosphere. For ease of display, all albedos except the one shown in dark green (second from the bottom) are displaced on the $y$-axis.}
\label{fig:albedo_resolution}
\end{figure}

Our reflection spectra were calculated on a uniform wavelength grid with 1 \AA\ spacing. However, Raman features could be detected in observations performed with coarser spectral resolution than that. In Fig. \ref{fig:albedo_resolution} we degrade our albedo spectra for the clear and cloudy ($P_\mathrm{max}=1$ bar) H$_2$/He atmospheres to spectral resolutions of $R=500$, $R=200$ and $R=50$. Most Raman scattering features, even the strongest ones corresponding to calcium H and K lines in the solar spectrum, are smeared out in the low-resolution (R=50) spectrum. They can be more easily discerned in the medium-resolution (R=200) spectrum, especially if the atmosphere is clear. However, this model is probably very optimistic and to identify these features in more realistic scenarios, higher spectral resolutions (such as $R=500$) would be preferred.

\subsubsection{Detecting Raman Features with High-contrast Instruments}
\label{sec:direct}

\begin{figure*}
\centering
\includegraphics[height=0.33\textheight]{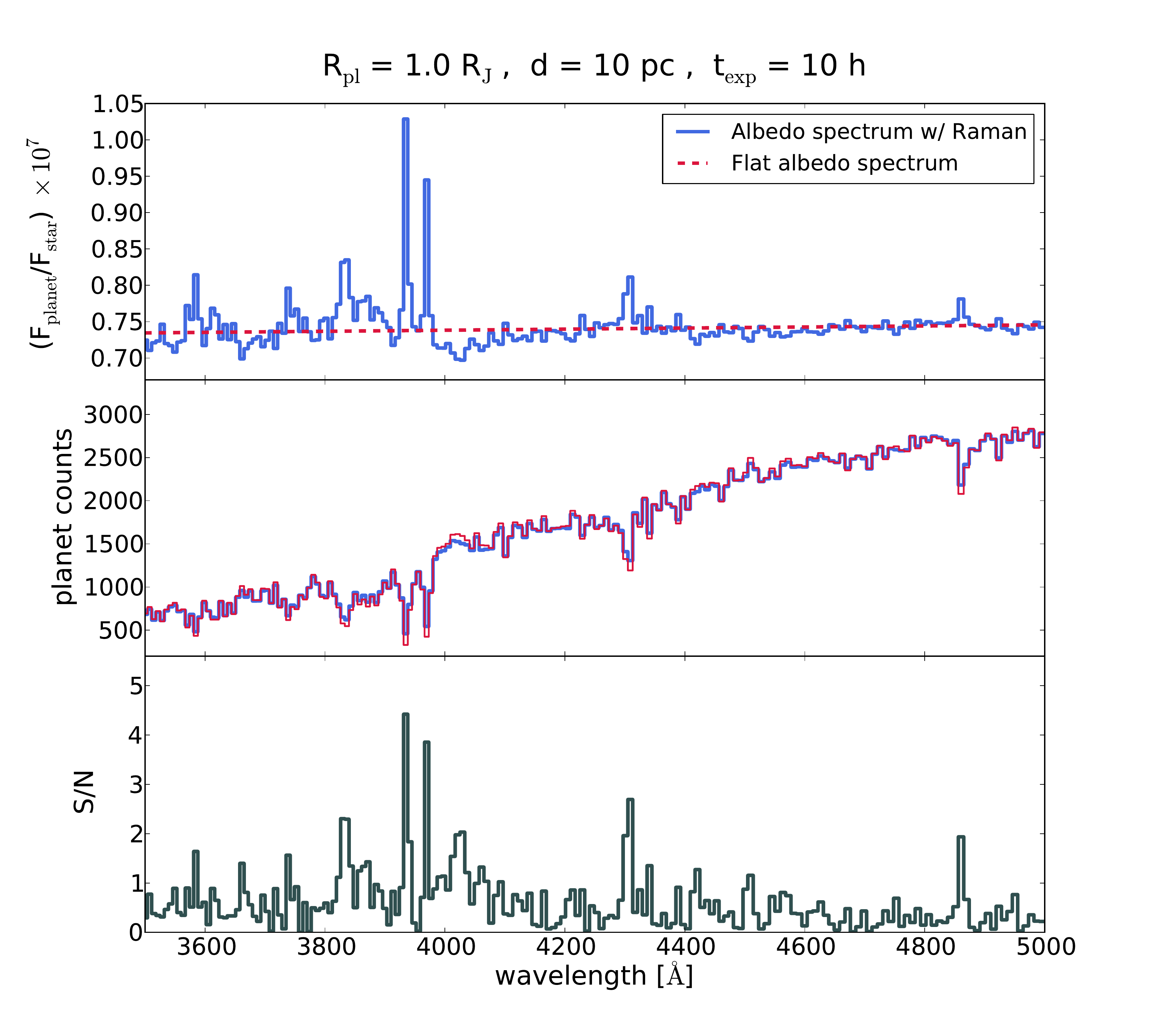}
\includegraphics[height=0.33\textheight]{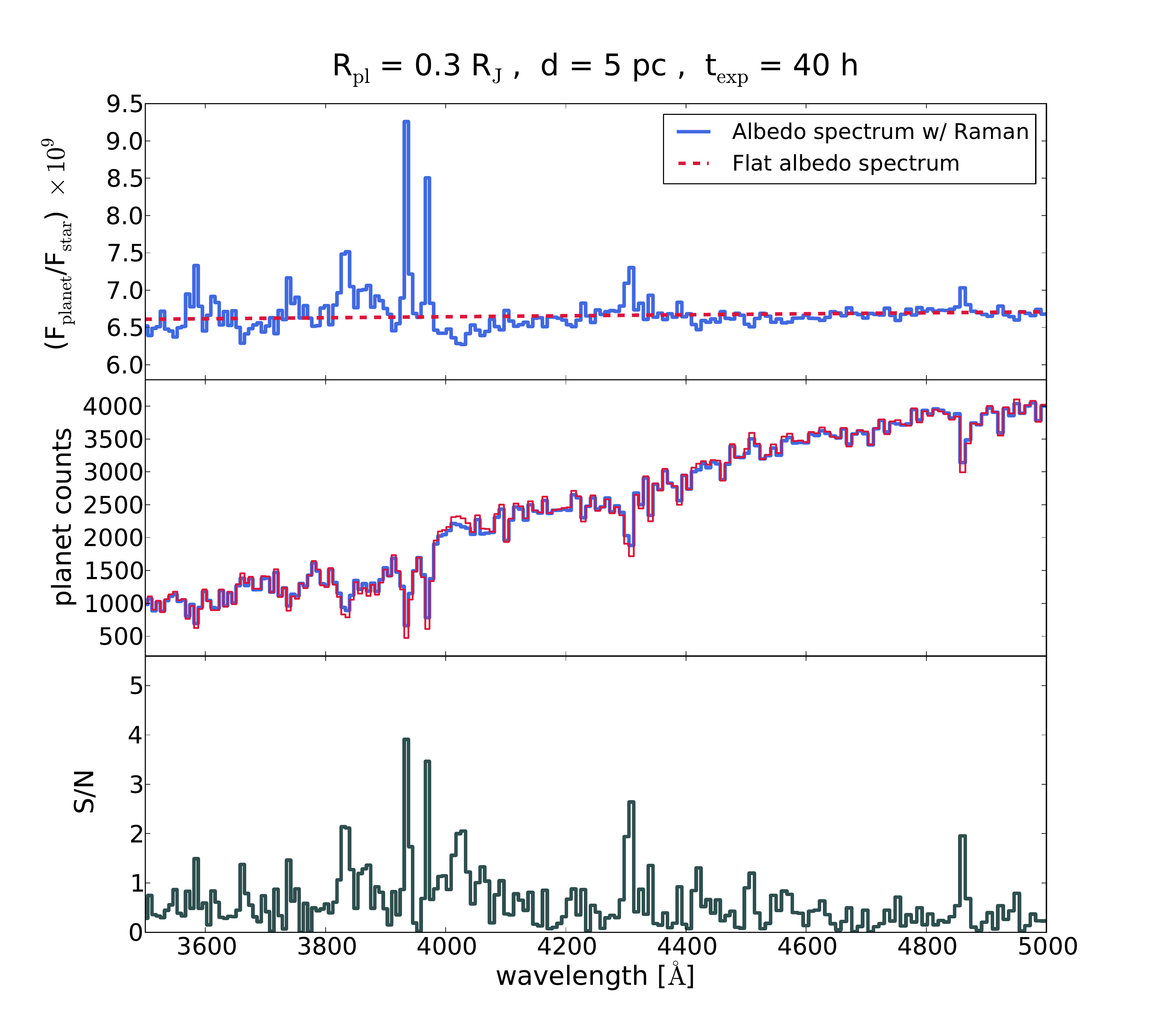}
\caption{Top panel: the planet-star flux ratio (PSFR) for a Jupiter- and a Neptune-sized planet (left and right, respectively) orbiting a Sun-like star at 0.8 AU separation, observed at the phase angle of $\alpha=80^\circ$, at the distance of 10 pc and 5 pc, respectively. The blue line shows the PSFR from a model atmosphere that includes the effects of Raman scattering, degraded to spectral resolution of $R=500$. The red dashed line shows a flat PSFR (i.e. without Raman features). Middle panel shows the total photon counts from the planet, observed with a 5-meter coronagraph-equipped telescope for the total exposure time of 10 and 40 hours. Bottom panel shows the ratio of signal to noise, at each spectral bin, to which signals with and without Raman features can be distinguished. Total S/N over all bins is $\sim 17$ in both cases.}
\label{fig:snr1}
\end{figure*}

In this section, we investigate the feasibility of observing Raman features using high-contrast imaging with a space telescope. Several approaches have been proposed, including both coronagraphs and starshades and using a range of telescope apertures. As an example in this initial study, we consider a 5 m telescope with a coronagraph achieving a contrast of $10^{-10}$. We use the instrument noise model described by \cite{Robinson2015} to make our predictions. Here we give a brief overview of the main expressions used in their noise model and the baseline set of instrument parameters.

The photon count rate of a planet at a distance $d$ from the observer is given by
\begin{equation}
c_p = \pi qf_{pa}\mathcal{T}\frac{\lambda^2}{hc\mathcal{R}}\left(\frac{D}{2}\right)^2F_{p,\lambda}(d)
\end{equation}
where $F_{p,\lambda}(d)$ is the planet specific flux density, related to the stellar flux and the geometric albedo of the planet though equation~(\ref{eq:pl_st_flratio}). Descriptions and values of other parameters are listed in Table~\ref{tab:parameters}.

\begin{table}
\centering
\label{tab:parameters}
\caption{Astrophysical and telescope parameters}
\begin{tabular}{ L  c  R  }
\hline
 \mbox{Symbol} &  Description & \mbox{Value} \\
 \hline
    R_\mathrm{pl} & planet radius (R$_{J}$) & 0.3\ /\ 1.0\\
   d & distance to observer (pc) & 5\ /\ 10\\
   a & planet-star separation (AU) & 0.8 \\
   \alpha &   phase angle (deg) & 80\\ 
   \Delta t_\mathrm{exp} &   exposure time (hours) & 40\ /\ 10\\ 
    \hline
   q & quantum efficiency & 0.9 \\
   f_{pa} & frac. of light in aperture & 0.87 \\
  \mathcal{T} & throughput & 0.05 \\
  \mathcal{C} &   coronagraph contrast & 10^{-10}\\
   \mathcal{R} & spectral resolution & 500 \\
   D & telescope diameter (m) & 5 \\
   \hline
\end{tabular}
\end{table}

The background count rate consists of contributions from zodiacal ($c_z$) and exozodiacal ($c_{ez}$) light, stellar light leaked through the coronagraph ($c_{lk}$), dark current ($c_{D}$) and read noise ($c_{R}$):
\begin{equation}
c_b = c_{z} + c_{ez} + c_{lk} +c_{D} + c_{R}  \ \mbox{.}
\end{equation}
We use the expressions and parameters from \cite{Robinson2015} to compute all these different contributions to the background count rate. For the total exposure time $\Delta t_\mathrm{exp}$, the total number of background counts is $C_b = c_b \Delta t_\mathrm{exp}$ and the total number of photons coming from the planet is $C_p = c_p \Delta t_\mathrm{exp}$. The total noise count is $C_{noise} = \sqrt{C_p + 2C_b}$.

We estimate the signal-to-noise-ratio (S/N) to which Raman features in the albedo spectrum could be detected and distinguished from a flat, featureless albedo spectrum of an atmosphere in which there was no significant Raman scattering, due to an opaque high-altitude cloud deck. Our `signal' then is the difference between the photon count coming from a planet whose albedo spectrum shows prominent Raman features ($C_{p,R}$, blue lines in Fig.~\ref{fig:snr1}) and the photon count coming from a planet whose albedo spectrum would be featureless ($C_{p,f}$, red dashed lines in Fig.~\ref{fig:snr1}). S/N is given by
\begin{equation}
S/N = \frac{\vert C_{p,R} -C_{p,f}\vert}{C_{noise}} = \frac{\vert c_{p,R}-c_{c,f}\vert }{\sqrt{c_{p,R} + 2c_b}}\sqrt{\Delta t_{exp}}
\end{equation}

In Fig.~\ref{fig:snr1} we show the planet-star flux ratio for a clear warm H$_2$/He atmosphere on a Jupiter- and Neptune-sized planets. Both planets are assumed to be orbiting Sun-like stars at a separation of 0.8 AU, observed at the phase angle of $\alpha=80^\circ$ (i.e. shortly before quadrature). Instead of just multiplying the calculated albedo with some specific planet phase function, we compute the disk integrated intensity at this phase by performing a new set of radiative transfer calculations that take into account different incident and reflected angles ($\mu_0$ and $\mu$) for $\alpha=80^\circ$, taken from \cite{HorakLittle1965}. The planets are assumed to be at a distance of 10 pc and 5 pc from Earth, for the larger and smaller planet respectively. We show the planet-star flux ratio observed on Earth and the photon counts assuming the planets are observed with a 5-meter coronagraph-equipped telescope, with spectral resolution $R=500$ and for the total exposure time of 10 hours for the larger and 40 hours for the smaller planet. The bottom panel shows S/N to which we can distinguish between Raman features and a flat albedo spectrum, for each wavelength bin. As expected, the highest S/N is achieved in wavelength bins corresponding to strong Raman peaks associated with the Ca II H and K lines. The cumulative S/N for all wavelength bins is $\sim 17$ in both cases.

\begin{figure}
\centering
\includegraphics[width=0.5\textwidth]{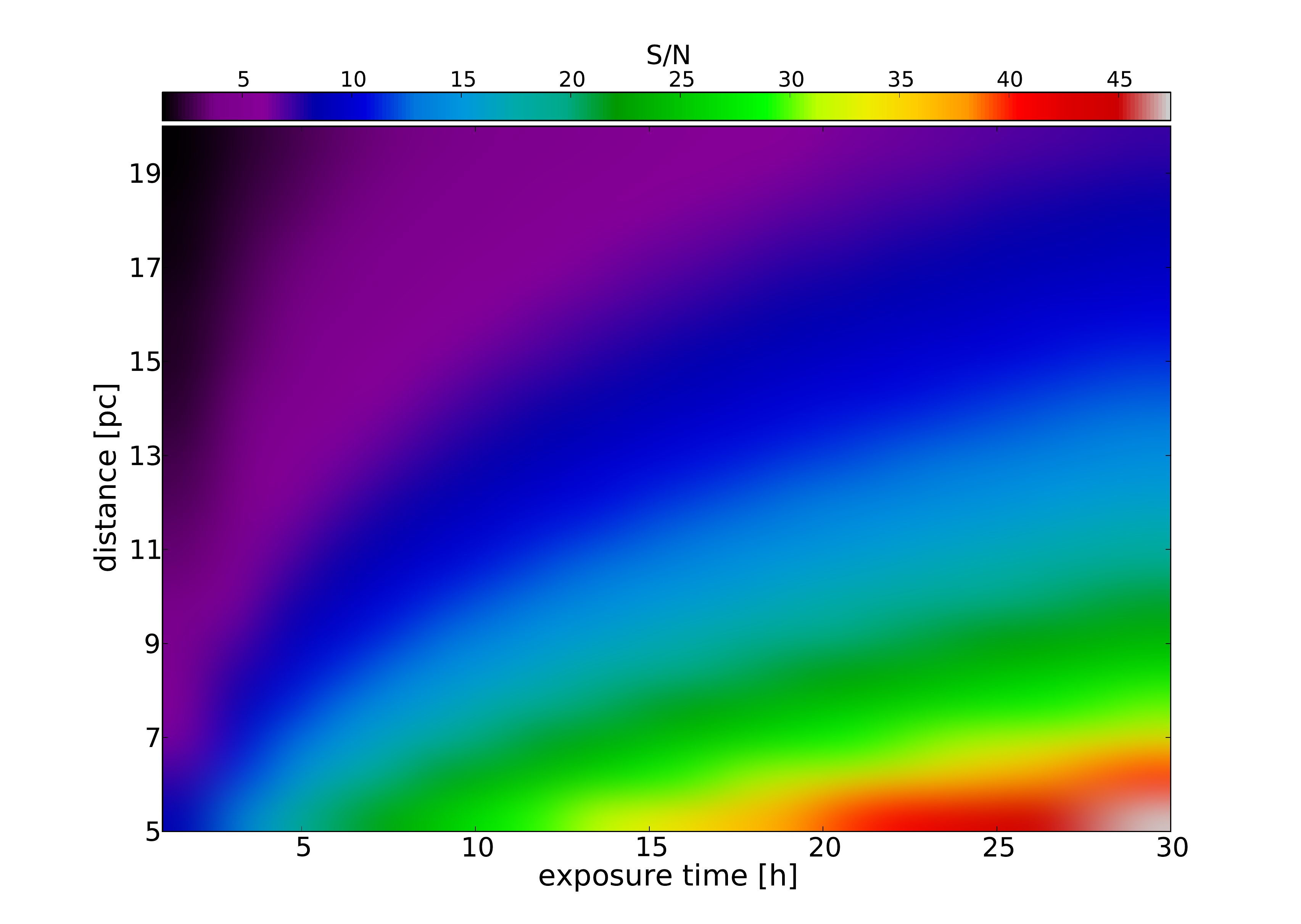}
\caption{Cumulative S/N over the wavelength range 3500-5000~\AA\ to which Raman features can be detected in a Jupiter-sized planet (orbiting the host star at 0.8 AU and observed near quadrature), depending on the distance to Earth and the total exposure time for observations with a 5-meter high-contrast ($10^{-10}$) telescope.}
\label{fig:snr1b}
\end{figure}

We perform the same type of analysis for a population of planets with radius $R= 1 R_J$, located at a range of distances from Earth (from 5~pc to 20~pc) and for observations with a range of total exposure times (1-30 hours; all other telescope/observational parameters are as in Table~\ref{tab:parameters}). We show the total S/N over the wavelength range 3500-5000~\AA\ in Fig.~\ref{fig:snr1b}. Our results suggest that detecting the signatures of Raman scattering in giant exoplanets could be feasible with a 5-meter coronagraph operating at these wavelengths. The analyzed distance range should contain several tens (or $\sim 100$) of solar-type stars.

\subsubsection{Detection Method II: Blended Light Case}
\label{sec:highres}

\begin{figure}
\centering
\includegraphics[width=0.5\textwidth]{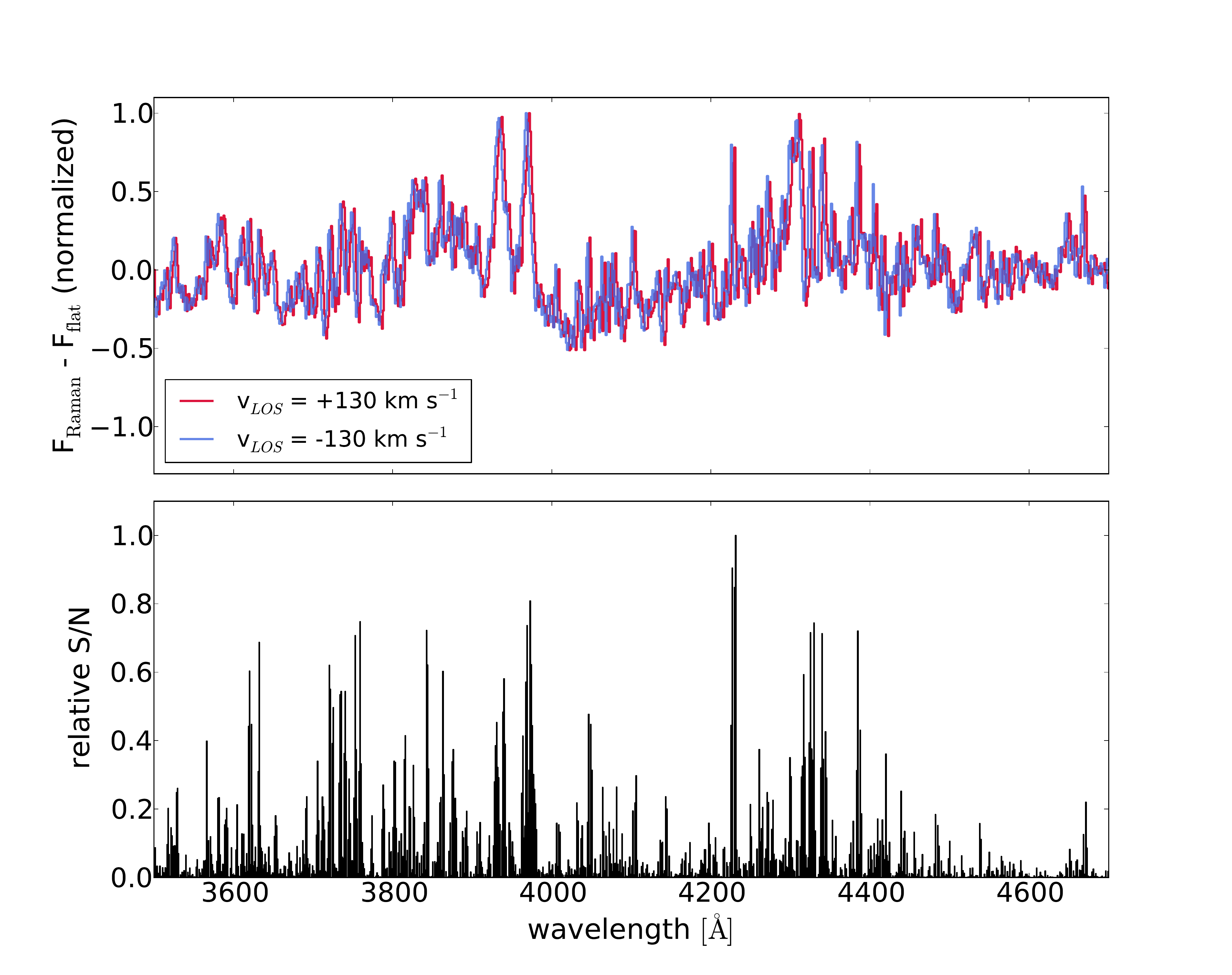}
\caption{Upper panel: Raman features in the reflected light get blue(red)shifted as a function of the line-of-sight velocity $v_{LOS}$ of the planet. Due to this property, the signal from the planet can be separated from the signal of its host star. This technique can be used to detect Raman scattering in the reflected light with high spectral resolution observations. The signal is dominated by wavelengths corresponding to strong and narrow lines, as shown by the normalized S/N distribution in the lower panel.}
\label{fig:red_blueshift}
\end{figure}

\begin{figure*}
\centering
\includegraphics[height=0.28\textheight]{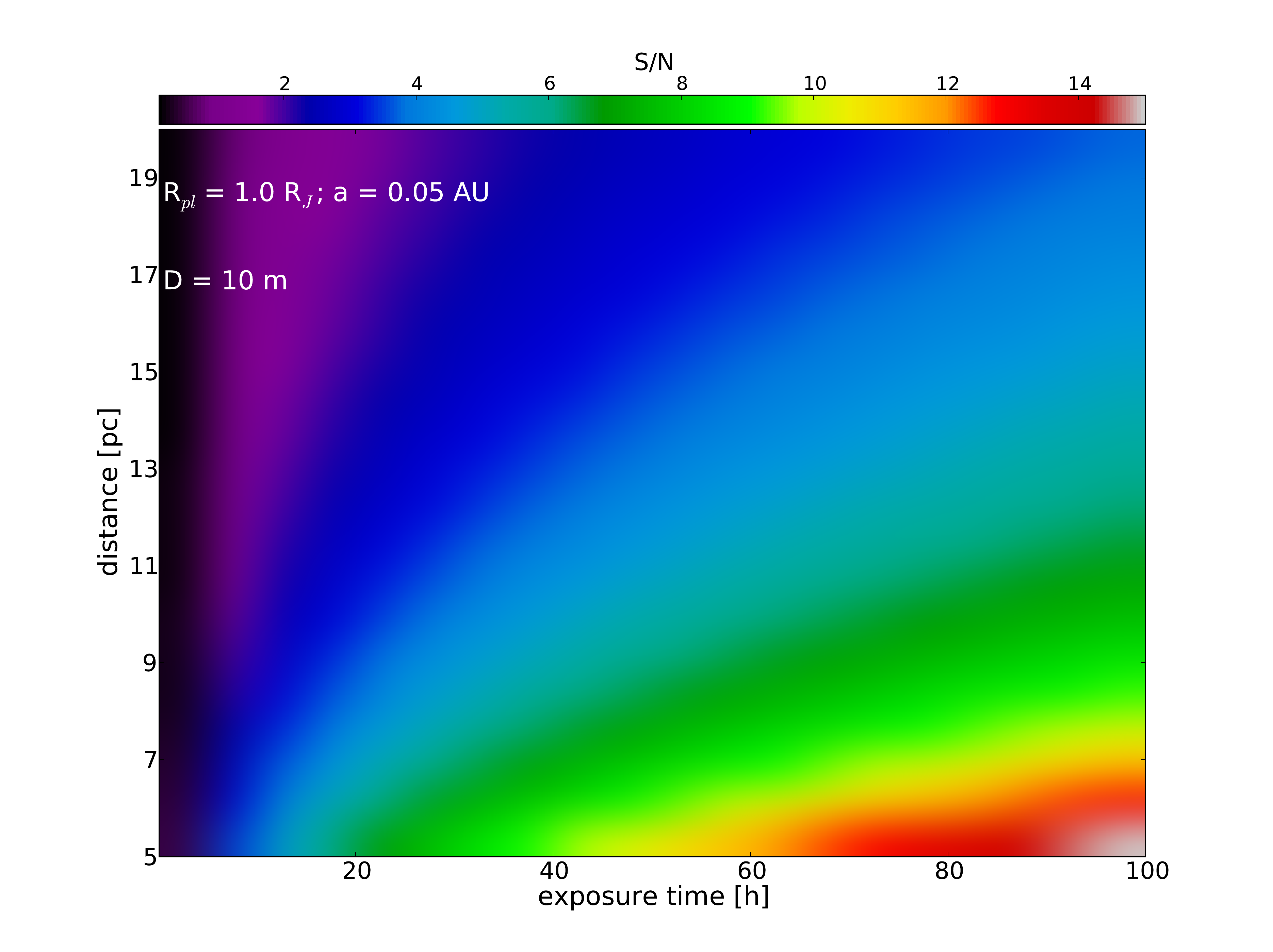}
\includegraphics[height=0.28\textheight]{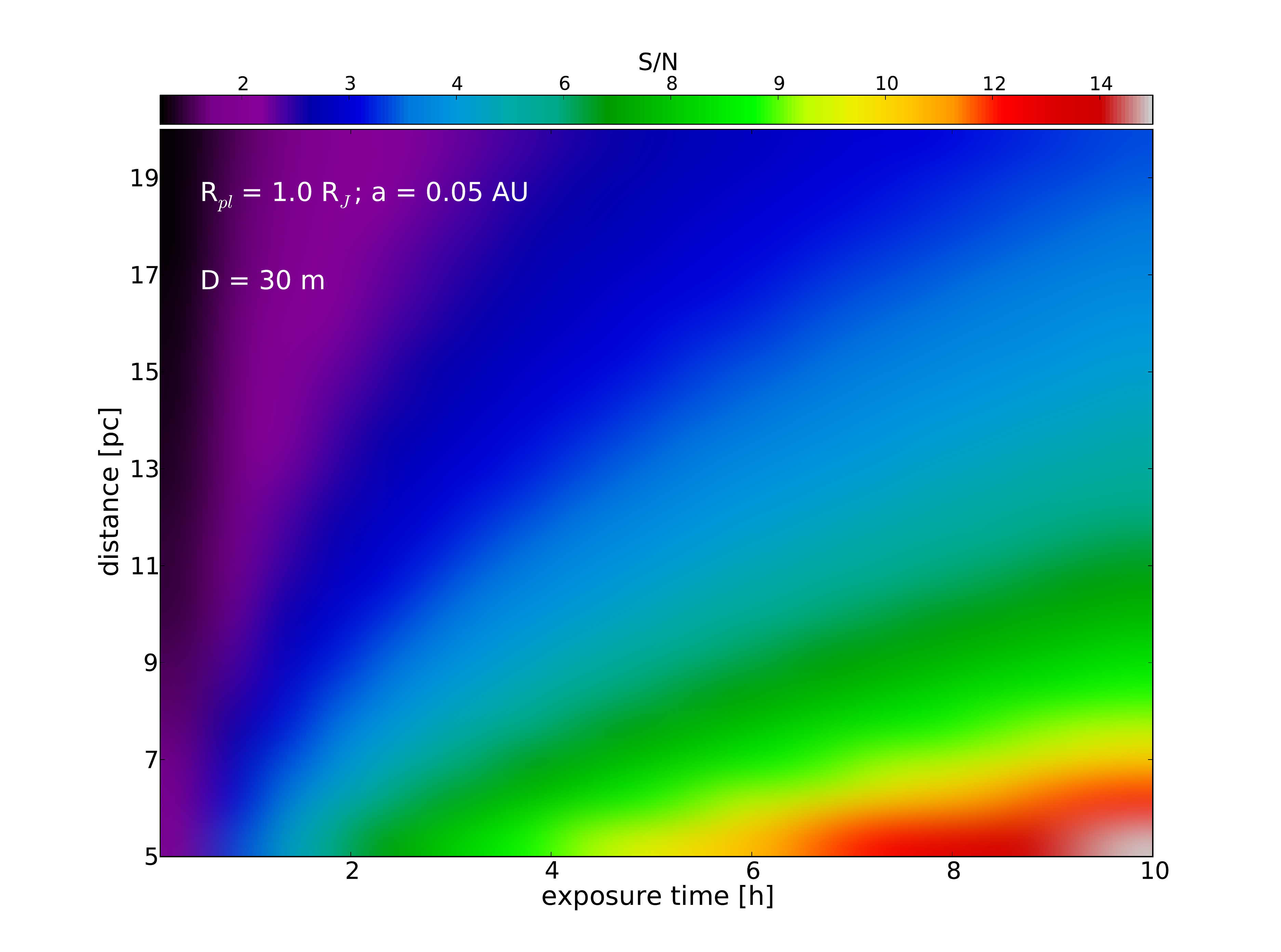}
\caption{Estimates of the total S/N for observations of a Jupiter-sized planet orbiting a Sun-like star at 0.05 AU separation, observed near quadrature ($\alpha=80^\circ$), for different combinations of distance to Earth and the total exposure time on a 10-meter (left panel) and a 30-meter ground-based telescope (right panel). Note the difference in the $x$-axis range in the two panels -- the time needed to reach a certain S/N for the same planet on the smaller-aperture telescope is about an order of magnitude greater than the exposure time needed on the larger telescope.}
\label{fig:snr2}
\end{figure*}

Next we consider observations in which the light from a planet is blended with the light from the host star. The principle is similar to observations using high-dispersion spectroscopy \citep[e.g.][]{Snellen2010,  Brogi2012,Birkby2013, Lockwood2014}. The total number of photons over a certain wavelength range reaching the observer on Earth from the planet-star system at the orbital phase $\Phi_i$ is given by $N_{s+p}(\Phi_i)$. At some other phase, $\Phi_j$, the total number of photons is given by a data vector $N_{s+p}(\Phi_j)$. We first look at the covariance matrix between the two vectors 
\begin{equation}
Cov(N_{s+p}(\Phi_i), N_{s+p}(\Phi_j)) \equiv C_{ij}= \delta_{ij} N_{s} \ \mbox{,}
\label{eq:cov_two}
\end{equation}
where $N_s$ is the number of photons coming from the star alone and $\delta_{ij}$ is the Kronecker symbol. 
The Fisher matrix is given by
\begin{equation}
F_{\alpha \beta} = \sum_{ij}[C^{-1}]_{ij} \frac{\partial N_{s+p}(\Phi_i)}{\partial p^\alpha} \frac{\partial N_{s+p}(\Phi_j)}{\partial p^\beta} \ \mbox{,}
\end{equation}
where $p$ denotes some set of parameters on which the covariance matrix does not depend. We find the inverse of the covariance matrix in equation~(\ref{eq:cov_two}) using the Taylor expansion
\begin{equation}
C^{-1}_{ij} = \frac{[N_{s+p}(\Phi_i) - N_{s+p}(\Phi_k)]^2}{2N_s} \ \mbox{.}
\label{eq:two_phase}
\end{equation}

If we generalize this procedure to having observations at $M$ different phases we get
\begin{eqnarray}
\nonumber C^{-1} &=& \frac{1}{N_s}\left(1- \frac{1}{M}\right) \sum_{i=1}^M N_{s+p}(\Phi_i)^2 \\
&-& \frac{1}{N_sM} \sum_{i=1}^M \sum_{\substack{j=1\\ j\neq i}}^M N_{s+p}(\Phi_i) N_{s+p}(\Phi_j)  \ \mbox{.}
\end{eqnarray}
Taking the square root of this, we get the signal-noise ratio for each spectral bin:
\begin{equation}
S/N = \sqrt{C^{-1}} \ \mbox{.}
\end{equation}

To get an estimate of how likely it would be to observe signatures of Raman scattering using this method, we use the simple two-phase formula (equation~\ref{eq:two_phase}) for the case of a Jupiter-sized planet orbiting a Sun-like star at a separation of 0.05 AU. We assume that the planet is observed near both quadratures ($\alpha=\pm 80^\circ$), where is has a radial velocity of $\pm 130$~km~s$^{-1}$ with respect to the star, and we impart the blue(red)shift to the planetary spectrum accordingly. We define our signal as the difference between the observed photon number count in the case of a planetary atmosphere with Raman albedo features and an atmosphere with a flat, featureless albedo. This is shown in Fig.~\ref{fig:red_blueshift} for the case of planet moving away from ($v=+130$~km~s$^{-1}$) and moving towards the observer ($v=-130$~km~s$^{-1}$). Raman features in the planetary spectrum get blue(red)shifted as the planet orbits the host star. The wavelength shift is only $\sim 2$~\AA, hence moderately high spectral resolution is needed to detect these subtle changes with orbital phase. In this case we use $R=3000$. The Doppler-shifting approach is more favorable for narrow spectral features, hence the contributions to the total S/N are more evenly distributed over the entire wavelength range than in the case of the previous method we considered, where most of the signal came from prominent (and fairly broad) Raman features associated with the stellar calcium lines. 

In Fig.~\ref{fig:snr2} we show the calculated S/N using the two-phase formula, i.e. the difference between the red and the blue curve in Fig.~\ref{fig:red_blueshift}. The S/N is evaluated over the 3500 - 5000 \AA\ wavelength range, for different combinations of distance from Earth and total exposure time on a 10-meter (left panel) or a 30-meter (right panel) ground-based telescope. Telescope throughput is assumed to be 15\%  over the analyzed wavelength range (in reality this value is not constant, it degrades towards the blue end of the spectrum). Our estimates suggest that detecting the signatures of Raman scattering in nearby exoplanets, using this method, would be very challenging with the currently existing telescopes. However, it should be possible with the next generation of large ground-based observatories if the photon-counting noise limit can be reached for this type of measurement.

\subsection{Caveats and Future Work}

In this study we focus our attention on atmospheric scattering, in particular on the effects of Raman scattering and its imprint in the geometric albedo of exoplanets. We do not include absorption from atomic or molecular species in our treatment of atmospheric radiative transfer. In an absorbing atmosphere the effects of scattering are less pronounced than in an atmosphere in which absorption is negligible. Absorption leads to an overall decrease in the geometric albedo, particularly at wavelengths corresponding to strong absorption lines, where albedo can be significantly diminished compared to out-of-line continuum. The most likely sources of absorption in the wavelength range that we analyze in this work ($\sim 300 - 500$ nm) are alkali atoms sodium and potassium, with absorption lines at 330 nm (Na) and 404 nm (K). The strength of these lines depends on the vertical profiles of the Na and K abundance in the form of free atoms. Atmospheric models used in this initial study are not complex enough to include these effects self-consistently.

Similarly to filling-in of \textit{stellar} absorption lines by the Raman effect, the absorption lines from \textit{atmospheric} absorbers also become less deep due to photons that are Raman-shifted into the lines. This effect could lead to underestimates in the inferred atmospheric abundances of absorbing species if Raman scattering is not properly treated. \cite{Sromovsky2005} found that this effect is present at a few percent level even in the near-IR, where Raman scattering is much weaker than at shorter wavelengths discussed in this work. According to their calculation, in a clear atmosphere of a Neptune-like planet, about 4\% of the light coming from deep methane bands in the near-IR is Raman-scattered light. We leave the analysis of filling-in of atmospheric absorption lines in the near-UV and the visible part of the spectrum for future work.

\section{Summary and Conclusions}
\label{sec:conclusion}

In this paper we investigate the effects of Raman scattering by H$_2$ and N$_2$ molecules in the atmospheres of exoplanets. This process leaves distinct signatures imprinted in the reflected light and the planetary geometric albedo that could, if detected, be used to probe the physical conditions and the chemical composition of the atmosphere. Here we summarize the basics of Raman scattering, our results and conclusions:
\begin{itemize}
\item Raman scattering, unlike Rayleigh scattering, changes the frequency of the scattered photons. The intensity of Raman-scattered light at some frequency $\nu$ depends on the intensity of the incident light at the Raman-shifted frequency $\nu + \Delta\nu$. The frequency change $\Delta\nu$ is determined by the energy-level structure of the scattering molecule, and therefore can be used as a spectroscopic signature of the molecule.
\item Raman-scattered photons fill-in absorption lines in the reflected spectrum that are caused by the lower radiation intensity at wavelengths corresponding to absorption lines in the incident spectrum. This causes sharp peaks in the albedo spectrum at the wavelengths of strong lines in the stellar spectrum, such as the calcium lines at $\lambda \sim 390$ nm in the solar spectrum. These albedo enhancements are the most prominent and most easily detectable Raman features in the albedo spectrum. Under favorable conditions of a deep and clear atmosphere, the albedo at these wavelengths can be enhanced by almost a factor of 2 compared to the surrounding continuum.
\item Albedo enhancements are strongest for clear atmospheres and atmospheres with deep (low-altitude) clouds, whereas high-altitude clouds diminish the effect of Raman scattering. Hence, the strength of Raman peaks in the planetary albedo can be used to constrain the presence and the altitude of thick clouds.
\item Raman scattering produces `ghosts' of absorption lines in the incident spectrum. They are seen as small dips in the albedo at frequencies that are shifted by a fixed amount $\Delta\nu$ (Raman shift) from the original line in the stellar spectrum. Individual ghost lines are weak (on the order of a few percent change in the albedo), however they are present at fixed distances from all stellar lines and thus might be detected statistically using cross-correlation techniques. By detecting the ghost lines and measuring their Raman shifts, the scattering molecule can be uniquely identified. This might be the most direct way to \textit{spectroscopically} confirm the presence of spectrally inactive molecules like H$_2$ or N$_2$ in the visible part of the spectrum.
\item The strength of individual ghost lines depends on the population of the corresponding initial molecular state and hence depends on the temperature of the medium. If the intensity of ghost lines produced by the same stellar line, but originating from different initial molecular states, could be measured, the temperature of the atmosphere could be inferred. 
\item The most prominent Raman features can be discerned even in spectra of moderate spectral resolution of $R =\lambda/\Delta\lambda$ of a few hundred. 
\item Detecting Raman features in nearby hot exoplanets may be possible with the next generation observational facilities, such as the 30-meter-class ground-based telescopes.
\item For exoplanets farther from their host stars, a promising method of detecting the signatures of Raman scattering would be using space-based coronagraph/starshade-equipped telescopes with spectrographs operating in the blue and near-UV wavelengths.
\end{itemize}

\acknowledgments
We thank the anonymous referee for their insightful comments which helped to improve the quality of the paper. We thank the developers of DISORT for making their code publicly available. AO thanks Kelly Chance for providing tables with Raman scattering cross sections for molecular nitrogen, used for comparison with our calculated cross sections. AO also thanks Dimitri Mawet and his group for useful discussions and feedback on an early version of this paper. AO and CMH are supported by the U.S. Department of Energy, the David \& Lucile Packard Foundation, and the Simons Foundation. KH acknowledges support from the Swiss National Science Foundation, the PlanetS NCCR and the MERAC Foundation.

\appendix
\section{Rayleigh and Raman Scattering Cross Sections}
\label{sec:cross_sections}

\subsection{Hydrogen}

Many previous studies of Raman scattering used hydrogen Raman and Rayleigh cross sections from \citet{FordBrowne1973}, which include transitions from initial molecular states with rotational quantum number $J\leq 3$. This is well suited for studies of cold atmospheres of giant planets in the Solar System. However, atmospheres of exoplanets can have much higher temperatures in which a significant fraction of molecular hydrogen populates states with $J>3$. Since we were unable to find Rayleigh and Raman cross sections for $J>3$ states in the literature, we calculate them using the formulae from \cite{Dalgarno1962}.

The cross sections defined in \citet{Dalgarno1962} measure the the total (angle-integrated) outgoing power (in erg s$^{-1}$) divided by the incident energy flux (in erg cm$^{-2}$ s$^{-1}$). We define cross section as the rate of scattering events (in units of events s$^{-1}$) divided by the incident photon flux (in photons cm$^{-2}$ s$^{-1}$), hence we need to multiply the cross sections given in \cite{Dalgarno1962} by the ratio of the outgoing ($\lambda$) and the incident ($\lambda^\prime$) wavelength. Because of this correction factor our cross sections have wavelength dependence proportional to $(\lambda^{\prime}\lambda^3)^{-1}$, instead of $(\lambda)^{-4}$.

For a transition between the initial molecular state characterized by rotational and vibrational quantum numbers $J_i$ and $v_i$, to the final state with quantum numbers $J_f$ and $v_f$, the cross section is given by
\begin{equation}
Q_S(v_i, J_i; v_f, J_f) = \frac{128 \pi^5}{9\lambda^{\prime}\lambda^3}\frac{(J_i+1)(J_i+2)}{(2J_i+3)(2J_i+1)} \vert \langle R_i\vert \gamma(r) \vert R_f\rangle \vert^2 \ 
\label{eq:csHS}
\end{equation}
in the case of the S-branch transitions ($J_f = J_i +2$);
\begin{equation}
Q_Q(v_i, J_i; v_f, J_f) = \frac{128 \pi^5}{9\lambda^{\prime}\lambda^3}\left[3 \vert \langle R_i \vert \alpha(r) \vert R_f\rangle \vert^2 +\frac{2}{3}\frac{J_i(J_i+1)}{(2J_i-1)(2J_i+3)}\vert \langle R_i\vert \gamma(r) \vert R_f \rangle \vert^2\right] \
\label{eq:csHQ}
\end{equation}
for the Q-branch transitions ($J_f = J_i$; Rayleigh scattering is a special case of Q-branch transitions, one for which $v_i = v_f$ and $\lambda^\prime = \lambda$); and
\begin{equation}
Q_O(v_i, J_i; v_f, J_f) = \frac{128 \pi^5}{9\lambda^{\prime}\lambda^3}\frac{J_i(J_i-1)}{(2J_i-1)(2J_i+1)} \vert \langle R_i\vert \gamma(r) \vert R_f\rangle \vert^2 \ 
\label{eq:csHO}
\end{equation}
for the O-branch ($J_f=J_i-2$). In the expressions above $\alpha$ and $\gamma$ are the average polarizability and the polarizability anisotropy of the molecule, given as functions of internuclear separation $r$. We use tabulated values of $\alpha(r)$ and $\gamma(r)$ from \cite{Rychlewski1980}. $R$ represents the radial wave functions of the H$_2$ molecule for the initial and final ro-vibrational state (denoted by $R_i$ and $R_f$, respectively) that are obtained by solving the Schr\"{o}dinger equation using the H$_2$ potential from \citet{Wolniewicz1993}. Our computations are performed in the limit of static polarizability, and are thus valid for photon energies small compared to the first electronic excitation energy of the H$_2$ molecule.

\begin{figure}
\centering
\includegraphics[height=0.26\textheight]{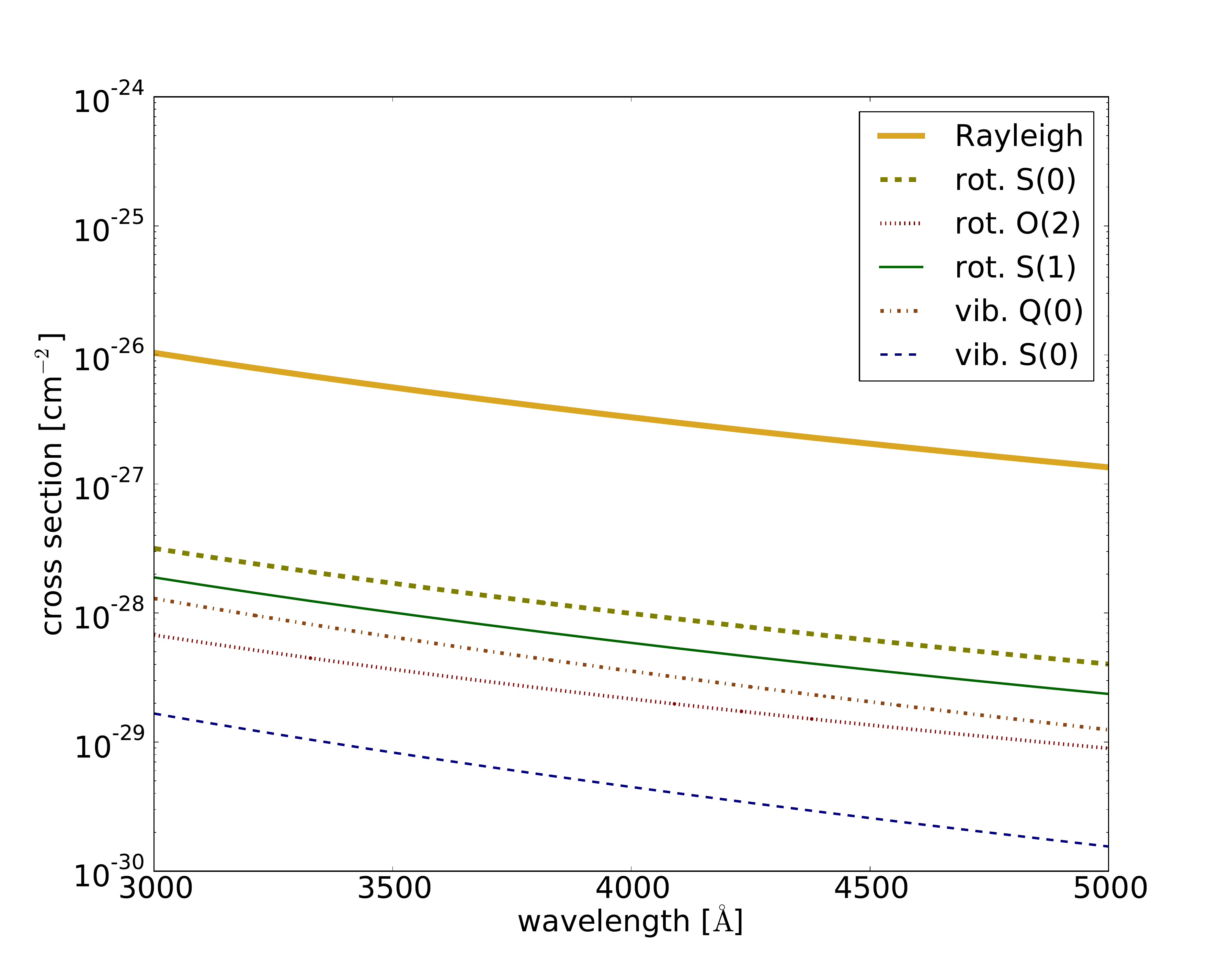}
\includegraphics[height=0.26\textheight]{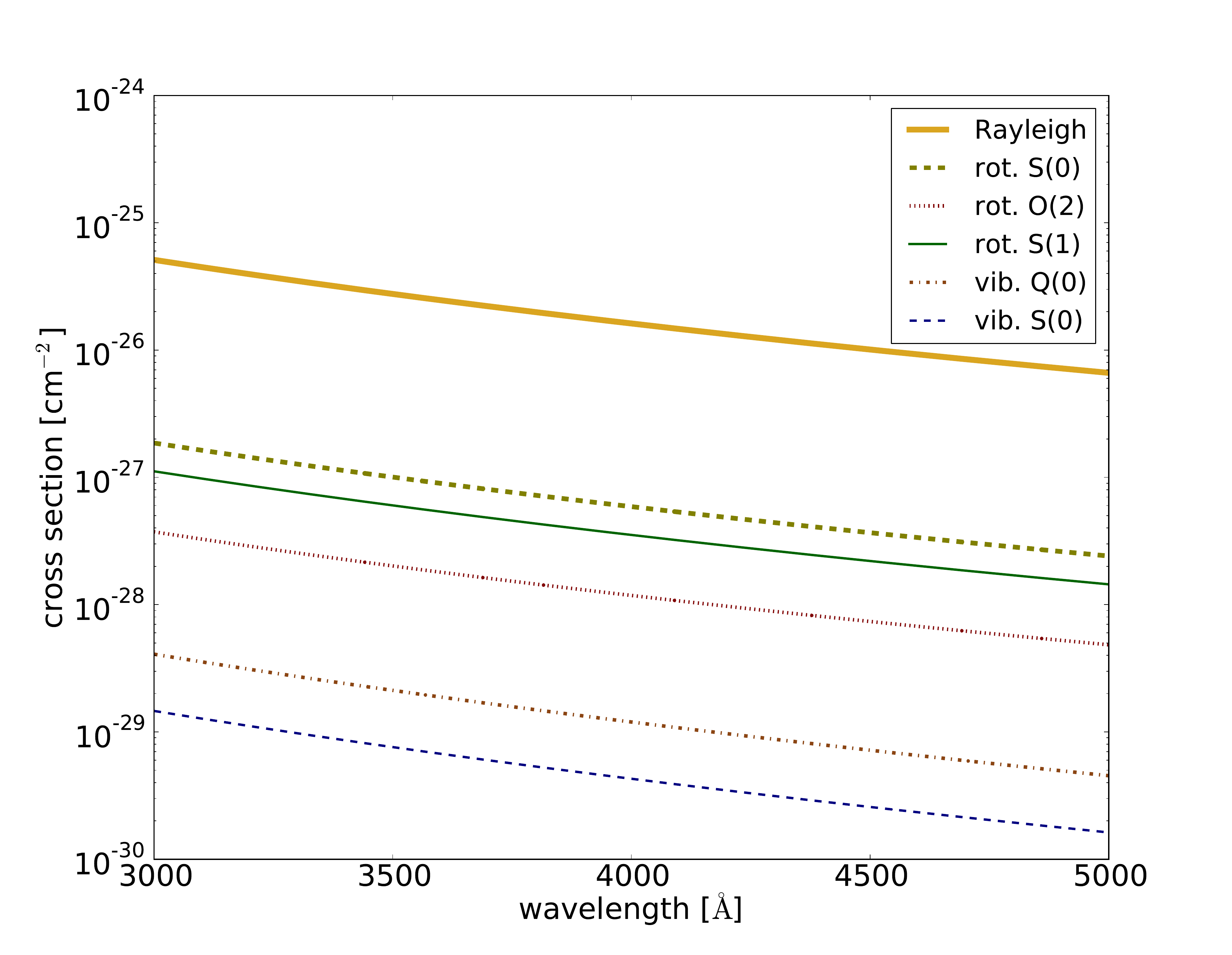}
\caption{Cross sections for Rayleigh scattering and several Raman scattering processes, for H$_2$ (left panel) and N$_2$ (right panel) in the wavelength range relevant for this study. Designations of Raman transitions have the following meaning: `vib' refers to vibrational transitions in which the vibrational quantum number of the molecule changes from 0 to 1 (ground to first excited state), `rot' describes pure rotational transitions in which the vibrational quantum number does not change. $S(i)$ denotes transitions in which the rotational quantum number increases by 2 (from $i$ to $i+2$), for $Q(i)$ the rotational quantum number does not change, and for $O(i)$ the rotational quantum number decreases by 2.}
\label{fig:cross_sections}
\end{figure}

The resulting cross sections are shown in Fig. \ref{fig:cross_sections} for Rayleigh scattering and a few different Raman transitions. Table \ref{tab:cross_sections} contains coefficients that can be used to evaluate the Rayleigh and Raman scattering cross sections. The cross section for a transition from an initial molecular state characterized by quantum numbers $v=0$ and $J_i$, to a final state with $v_f$ and $J_f$ is given by
\begin{equation}
Q(0,J_i; v_f, J_f; \lambda) = \frac{C}{\lambda^{\prime}\lambda^3} \ \mbox{[cm$^2$],}
\label{eq:poly_exp}
\end{equation}
where $\lambda$ and $\lambda^\prime$ are expressed in cm. The incident and scattered wavelengths are related by $\lambda^{\prime -1} = \lambda^{-1} + \Delta \nu$, where $\Delta \nu$ is the Raman shift of the transition in question.

Our values of cross sections for the most important transitions (Rayleigh, pure rotational Raman transitions and vibrational Q-branch transitions) agree with values of \citet{FordBrowne1973} to within $\sim 20 \%$ at the shortest wavelengths (around 300 nm) and within $\sim 5\%$ around 500 nm. The source of discrepancy is most likely the fact that we do not include the effects of dynamic polarizability, which become more relevant at higher energies. The differences between our cross sections and the results of \citet{FordBrowne1973} for vibrational S- and O-branch transitions are much larger (between 50-100\% difference). We were not able to establish the origin of this discrepancy. However, because these transitions are typically between one and two orders of magnitude weaker in cross section than the other Raman transitions (pure rotational and vibrational Q-branch), they do not significantly affect the main results of this work.

\begin{deluxetable}{C C C  R C | C C C  R C}  
\tablecolumns{13}
\tablecaption{Raman scattering cross sections for H$_2$}
\tablehead{   
  \colhead{$J_i$} &
  \colhead{$J_f$} &
  \colhead{$v_f$} &
  \colhead{$C$} &
  \colhead{$\Delta \nu$[cm$^{-1}$] }  &
    \colhead{$J_i$} &
  \colhead{$J_f$} &
  \colhead{$v_f$} &
  \colhead{$C$} &
  \colhead{$\Delta \nu$[cm$^{-1}$]} 
}
\startdata
0 & 0& 0 &8.40817110301\times10^{-45} &0.0 &1 &3 &1 &1.14413862330\times10^{-47} &4715.3283129\\
0 &0 &1 &1.56701934093\times10^{-46} &4163.3085901 & 2 &2 &0 &8.57452212207\times10^{-45} &0.0\\
0 &2 &0 &2.64973960879\times10^{-46} &354.5691479 & 2 &2 &1 &1.64943280591\times10^{-46} &4145.5863876\\
0 &2 &1 &2.08231744526\times10^{-47} &4500.1555355 & 2 &4 &0 &1.41322718574\times10^{-46}& 814.8723577\\
1 &1 &0 &8.54368656301\times10^{-45} &0.0 & 2 &4 &1 &8.99497252605\times10^{-48}& 4919.5253956\\
1 &1 &1 &1.66712462583\times10^{-46} &4157.3872895 & 2 &0 &0 &5.29947921758\times10^{-47} &-354.5691479\\
1 &3 &0 &1.61326313808\times10^{-46} &587.3565875 & 2 &0 &1 &5.47400061544\times10^{-48} &3808.7394422
\enddata
\label{tab:cross_sections}
\tablecomments{Coefficients for the scattering cross sections for H$_2$. The cross section for each transition is calculated according to equation~(\ref{eq:poly_exp}). The column $\Delta \nu$ contains the value of the Raman shift for each transitions. The table is available in its entirety (including cross sections for initial states up to $J_i=9$) in the machine-readable format. This portion is shown here for guidance purposes. }
\end{deluxetable}

\subsection{Nitrogen}

For the case of molecular nitrogen, we use the approximate formulae from \cite{Dalgarno1962} that rely on expanding the molecular polarizability and polarizability anisotropy around the equilibrium separation of nuclei $r_0$ according to
\begin{eqnarray}
\alpha(r) &=& \alpha(r_0) + (r-r_0)\alpha^\prime (r_0) +... \\
\gamma(r) &=& \gamma(r_0) + (r-r_0)\gamma^\prime (r_0) +... 
\end{eqnarray}
and on approximating the vibrational wave function of the molecule by a harmonic oscillator eigenfunction with an eigenvalue $h\nu_e(v+\frac{1}{2})$. In the equations above, $\alpha^\prime (r_0)$ and $\gamma^\prime (r_0)$ are the derivatives of the polarizability and polarizability anisotropy with respect to internuclear separation, evaluated at the equilibrium distance. 

The approximate formulae are:
\begin{itemize}
\item for Rayleigh scattering ($v_i=v_f$, $J_i=J_f$):
\begin{equation}
Q(v_i, J_i; v_f, J_f) = \frac{128\pi^5}{9\lambda^4}\left[3\alpha(r_0)^2 +\frac{2}{3}\frac{J_i(J_i+1)}{(2J_i-1)(2J_i+3)} \gamma(r_0)^2\right] \ \mbox{,}
\end{equation}
\item for pure rotational S-branch transitions ($v_f = v_i$, $J_f=J_i+2$):
\begin{equation}
Q(v_i, J_i; v_f, J_f) = \frac{128\pi^5}{9\lambda^{\prime}\lambda^3}\frac{(J_i+1)(J_i+2)}{(2J_i+3)(2J_i+1)}\gamma(r_0)^2 \ \mbox{,}
\end{equation}
\item for pure rotational O-branch transitions ($v_f = v_i$, $J_f=J_i-2$):
\begin{equation}
Q(v_i, J_i; v_f, J_f) = \frac{128\pi^5}{9\lambda^{\prime}\lambda^3}\frac{J_i(J_i-1)}{(2J_i-1)(2J_i+1)}\gamma(r_0)^2\ \mbox{,}
\end{equation}
\item for vibrational Q-branch transitions ($v_f = v_i+1$, $J_f=J_i$):
\begin{equation}
Q(v_i, J_i; v_f, J_f) = \frac{128\pi^5}{9\lambda^{\prime}\lambda^3}\left[3\alpha^\prime(r_0)+\frac{2}{3}\frac{J_i(J_i+1)}{(2J_i-1)(2J_i+3)}\gamma^\prime(r_0)^2 \right]\frac{(v_i+1)h}{8\pi^2\mu\nu_e} \ \mbox{,}
\end{equation}
\item for vibrational S-branch transitions ($v_f = v_i+1$, $J_f=J_i+2$):
\begin{equation}
Q(v_i, J_i; v_f, J_f) = \frac{128\pi^5}{9\lambda^{\prime}\lambda^3}\frac{(J_i+1)(J_i+2)}{(2J_i+3)(2J_i+1)}\gamma^\prime(r_0)^2 \frac{(v_i+1)h}{8\pi^2\mu\nu_e} \ \mbox{,}
\end{equation}
\item for vibrational O-branch transitions ($v_f = v_i+1$, $J_f=J_i-2$):
\begin{equation}
Q(v_i, J_i; v_f, J_f) = \frac{128\pi^5}{9\lambda^{\prime}\lambda^3}\frac{J_i(J_i-1)}{(2J_i-1)(2J_i+1)}\gamma^\prime(r_0)^2 \frac{(v_i+1)h}{8\pi^2\mu\nu_e} \ \mbox{,}
\end{equation}
\end{itemize}
where $\mu$ is the reduced mass of the molecule. For molecular nitrogen, we use the following set of parameter values: $\mu = 1.16\times 10^{-23}$ g, $\nu_e = 6.988 \times 10^{13}$ Hz, $\alpha(r_0) = 1.778 \times 10^{-24}$ cm$^{3}$, $\gamma(r_0) = 7.14 \times 10^{-25}$ cm$^3$, $\alpha^\prime(r_0) = 1.852 \times 10^{-16}$ cm$^2$, $\gamma^\prime (r_0) = 2.185 \times 10^{-16}$ cm$^2$ \citep{Bernath2005,Pecul2002}. Our cross sections agree to within $\lesssim 10\%$ with those used in studies of Raman scattering in the Earth's atmosphere (i.e. the Ring effect) by \citet{Chance1997}. 

\subsection{Helium}

For Rayleigh scattering on helium atoms we use the cross sections from \cite{Alipour2015}.

\section{Quadrature Coefficients}
\label{sec:quadrature}

To calculate the reflected light from a planet, we need to properly integrate over the surface of the planetary disk, as described in Section \ref{sec:albedo_calc}. In order to do that, we use the quadrature coefficients for order $n=6$ and the corresponding angles of incidence for a planet at full phase (phase angle equal to zero ) from \cite{HorakLittle1965}. Table \ref{tab:quad} gives the coefficients of the quadrature expansion ($a$ and $b$), along with the coordinates ($\psi$ and $\xi$) of the nine points on the disk used to compute the reflected light and the corresponding cosines of the incident angles ($\mu$). At full phase, the contributions from all four quadrants of the disk are the same, hence we can simply multiply the emerging intensity from the first quadrant by 4 to get the intensity from the entire planet.

\begin{deluxetable}{ c  c  c  c c  c  c c  }  
\tablecolumns{8}
\tablecaption{Quadrature coordinates and coefficients}
\tablehead{   
  \colhead{number} &
  \colhead{latitude ($\pi/2-\eta$)} &
  \colhead{longitude ($\zeta$)} &
  \colhead {$\psi$} &
  \colhead{ $\xi$} &
  \colhead{$a$} &
  \colhead{ $b$} &
  \colhead{ $\mu_0=\mu$} 
}
\startdata
    1 & 0.2244 (12.86$^\circ$) & 0.2409 (13.81$^\circ$) & 0.222521 & 0.238619 & 0.426576 & 0.467914 & 0.94676541 \\ 
    2 & 0.2244 (12.86$^\circ$) & 0.7224 (41.39$^\circ$) & 0.222521 & 0.661209 & 0.426576 & 0.360762 & 0.73139233 \\ 
    3 & 0.2244 (12.86$^\circ$) & 1.2011 (68.82$^\circ$)  & 0.222521 & 0.932470 & 0.426576 & 0.171324 & 0.35219154 \\ 
    4  & 0.6732 (38.57$^\circ$) & 0.2409 (13.81$^\circ$) & 0.623490 & 0.238619 & 0.274333 & 0.467914 & 0.75924690 \\ 
    5  & 0.6732 (38.57$^\circ$) & 0.7224 (41.39$^\circ$)  & 0.623490 & 0.661209 & 0.274333 & 0.360762 & 0.58653110 \\ 
    6  & 0.6732 (38.57$^\circ$) & 1.2011 (68.82$^\circ$) & 0.623490 & 0.932470 & 0.274333 & 0.171324 & 0.28243569 \\ 
    7 & 1.1221 (64.29$^\circ$) & 0.2409 (13.81$^\circ$) & 0.900969 & 0.238619 & 0.0844887 & 0.467914 & 0.42135023 \\
    8  & 1.1221 (64.29$^\circ$) & 0.7224 (41.39$^\circ$)  & 0.900969 & 0.661209 & 0.0844887 & 0.360762 & 0.32550020 \\ 
    9  & 1.1221 (64.29$^\circ$) & 1.2011 (68.82$^\circ$) & 0.900969 & 0.932470 & 0.0844887 & 0.171324 & 0.15673998 
\enddata
\label{tab:quad}
\tablecomments{The quadrature coordinates (latitude, longitude) or ($\psi$, $\xi$) and weights ($a, b$) used in calculating the integrated disk intensity according to equation~(\ref{eq:quad}). The last column ($\mu_0, \mu$) gives the cosine of the angle at which the radiation beam illuminates the atmosphere, and at which the light is reflected toward the observer (these two angles are equal at full phase). }
\end{deluxetable}

\bibliographystyle{yahapj}
\bibliography{refs_raman}

\end{document}